\documentclass[10pt,leqno]{amsart}
\usepackage{graphicx}
\usepackage{indentfirst,csquotes}
\usepackage{amssymb,amsthm,amsmath}
\usepackage{xcolor,paralist,hyperref,titlesec,fancyhdr,etoolbox}
\usepackage{lipsum}
\usepackage{natbib}
\usepackage{amsmath}
\usepackage[nopatch]{microtype}
\usepackage{booktabs}
\usepackage{graphicx}
\usepackage{multirow}
\usepackage{float}
\usepackage{placeins}
\usepackage[utf8]{inputenc}
\usepackage{textgreek}
\usepackage{booktabs}
\usepackage{multirow}
\usepackage{colortbl}
\usepackage{xcolor}
\usepackage{subcaption}
\usepackage{dashrule}
\usepackage{booktabs}
\usepackage{tabularx}
\usepackage{geometry}
\usepackage{setspace}
\usepackage{lineno}
\usepackage{tabularx}
\usepackage{caption}
\newcolumntype{C}{>{\centering\arraybackslash}X}

\titleformat{\section}{\normalfont\large\bfseries\centering}{\thesection}{1em}{}
\titlespacing*{\section}{0pt}{2ex}{1ex}
\titleformat{\subsection}{\normalfont\normalsize\bfseries}{\thesubsection}{1em}{}
\titlespacing*{\subsection}{0pt}{1.5ex}{0.5ex}
\titleformat{\subsubsection}{\normalfont\normalsize\itshape}{\thesubsubsection}{1em}{}
\titlespacing*{\subsubsection}{0pt}{1ex}{0.5ex}
\hypersetup{colorlinks=true, linkcolor=black, filecolor=black, urlcolor=black}

\begin{document}

\title{Uncovering Latent Pathological Signatures in Pulmonary CT via Cross-Window Knowledge Distillation}

\author[B.~Peng]{Bo Peng$^{*}$}
\author[W.~Xu]{Wujian Xu$^{*}$}
\author[K.~Wang]{Kun Wang}
\author[X.~Liao]{Ximing Liao}
\author[N.~Wang]{Na Wang}
\author[D.~Shi]{Daqian Shi}
\author[T.~Li]{Tian Li}
\author[J.~Gao]{Jing Gao}
\author[J.~Thygesen]{Johan Thygesen}
\author[Y.~Ji]{Yingqun Ji}
\author[H.~Wu]{Honghan Wu$^{\dagger}$}

\address{Bo Peng, Johan Thygesen, Honghan Wu: Institute of Health Informatics, University College London, London, UK}
\address{Kun Wang, Ximing Liao, Na Wang, Tian Li, Jing Gao, Yingqun Ji, Wujian Xu: Department of Pulmonary and Critical Care Medicine, Shanghai East Hospital, School of Medicine, Tongji University, Shanghai, China}
\address{Daqian Shi: Queen Mary University of London, London, UK}
\address{Honghan Wu: School of Health and Wellbeing, University of Glasgow, Glasgow, UK}

\email{honghan.wu@glasgow.ac.uk}
\date{\today}

\maketitle

\let\thefootnote\relax
\footnotetext{$^{*}$ These authors contributed equally to this work.}
\footnotetext{$^{\dagger}$ Corresponding author: \texttt{honghan.wu@glasgow.ac.uk}}

\begin{abstract}

\noindent\textbf{Purpose:} Multi-window CT imaging captures complementary pathological information from anatomical structures of differing densities. Existing deep learning approaches analyse windows independently and fuse representations only at later stages, failing to capture pathological nuances arising from systemic interplay between regions of differing densities. This study developed a cross-window knowledge distillation framework that learns latent feature representations.

\noindent\textbf{Materials and Methods:} The method was retrospectively evaluated on three cohorts: COPD-CT-DF (n=719), an RSNA PE subset (n=1,433), and an in-house CTPA dataset (n=161) for chronic thromboembolic pulmonary disease (CTEPD). Five CT windows were pre-computed per scan. A teacher encoder was identified through performance benchmarks on individual windows; student encoders were then trained via cross-window knowledge distillation to internalise the teacher's clinical priors. Mechanistic insights were validated through a clinician survey, evaluating how the model captures latent, cross-window anatomical and pathological signatures that resolve diagnostic ambiguity.

\noindent\textbf{Results:} On COPD-CT-DF, cross-window distillation significantly improved students' AUCs by 10.1-16.5 percentage points over baselines (all $P<.001$), from 0.75-0.81 to 0.90-0.94. An ensemble of the teacher and distilled students achieved accuracy 0.94 and AUC 0.9960 (vs. 0.77 and 0.8936 baseline). Distilled models corrected 21-24 misclassifications per window, capturing pathological signatures invisible to students. On the RSNA PE dataset, per-window AUC improved from 0.80-0.83 to 0.90-0.92. For CTEPD, fine-tuned adaptation achieved AUC 0.7481 vs. 0.6264 for supervised pretraining.

\noindent\textbf{Conclusion:} Cross-window knowledge distillation internalises pathological information invisible to traditional supervised approaches. By resolving feature interference and facilitating integrated pathological reasoning, it provides a generalisable approach for multi-window medical imaging analysis.
\end{abstract}

\section{Introduction}

Deep learning in medical image analysis relies heavily on effective data representation. Unlike natural images, information extraction from medical images, particularly computed tomography (CT) scans, is highly dependent on the choice of imaging parameters \citep{wu2024ai}. CT window settings are fundamentally a feature selection process: by controlling the mapping from Hounsfield Units (HU) to grayscale intensities, different windows extract tissue information from distinct density ranges within the same raw data. Lung windows (level: $-600$ HU) enhance low-density regions to visualize emphysema and airways, mediastinal windows (level: 20 HU) highlight vascular and soft-tissue structures, and bone windows (level: 250 HU) display osseous details \citep{webb2015fundamentals}. These window settings correspond to different pathological dimensions of disease: the manifestations of a single disease are often distributed across distinct density ranges, as exemplified by chronic obstructive pulmonary disease (COPD) and pulmonary embolism (PE). The former simultaneously involves parenchymal, vascular, and thoracic alterations, while the latter requires the assessment of intravascular filling defects alongside secondary parenchymal and pleural findings \citep{yang2021ct, celli2015official}. The radiologist's diagnostic process is precisely the synthesis of this complementary information, yet this cognitive workflow has not been adequately modeled in existing deep learning approaches.

Existing deep learning approaches to multi-window information utilization follow two main patterns, neither fully exploiting inter-window complementarity. The first category employs fixed window settings or HU value ranges as static parameters determined during preprocessing. For instance, Hu et al.'s pulmonary embolism detection model uses a fixed HU range \citep{hu2025high}, Yin et al.'s COVID-19 identification system \citep{yin2024automated} preprocesses to lung windows. This design rests on an implicit assumption: that some ``optimal'' window can sufficiently express disease features, thereby neglecting the complementary value of other windows. The second category attempts to fuse multi-window information, but primarily through input-level multi-channel stacking. For example, Hemalakshmi et al.'s PE-Ynet model \citep{hemalakshmi2024pe} uses lung, mediastinal, and bone windows as three-channel CNN input for pulmonary embolism detection, analogous to processing RGB images. While this shallow fusion strategy allows models to access multi-window information, it treats window integration as a data representation problem rather than a knowledge integration problem. By mixing disparate voxel densities at the network's lowest level, these models suffer from feature interference, limiting each window's ability to independently develop deep semantic representations.

The more fundamental issue lies in these approaches' understanding of window information, overlooking a critical fact: different windows correspond not to different observations of the same pathological process, but to orthogonal dimensions of the disease phenotype space. Parenchymal changes, vascular abnormalities, and structural deformations represent relatively independent pathological processes, each most clearly visible in different windows. Ideal multi-window learning should allow each window to develop its specific feature extractor, then integrate these heterogeneous features at higher semantic levels. Existing methods either abandon this complexity through single-window analysis or force premature fusion (i.e., before representation learning), failing to capture the systemic interplay between anatomical structures.

Knowledge Distillation (KD) provides a transformative pathway to address this challenge. While traditionally used for model compression~\citep{gou2021knowledge}, the core mechanism of distillation reveals possibilities for cross-representation knowledge transfer~\citep{wang2023learnable, huo2024c2kd}. When teacher and student process different input representations, distillation forces the student to internalise the teacher's high-level clinical priors rather than raw pixel information. This creates ``cross-window intuition'' where a student model (e.g., on Lung window) can be trained to identify latent pathological signatures - features that are effectively invisible to traditional models or human observers because they serve as proxies for systemic pathology usually restricted to the teacher's domain (e.g. on Mediastinal window). By operating in the feature space, KD achieves deep anatomical reasoning, allowing the model to resolve diagnostic ambiguity through integrated expertise.

\begin{table*}[!htbp]
\centering
\small
\caption{Characteristics of the three cohorts used in this study. Continuous variables are presented as mean $\pm$ standard deviation; categorical variables as n (\%). For all three datasets, the positive class corresponds to the disease or adverse outcome: Chronic obstructive pulmonary disease (COPD), acute Pulmonary embolism (PE), and progression to Chronic thromboembolic pulmonary disease (CTEPD), respectively.}
\label{tab:dataset_characteristics}

\begin{subtable}[t]{\linewidth}
\centering
\caption{Three cohorts for three diagnostic tasks. COPD-CT-DF was used as the derivation and internal validation data for COPD classification models. A subset of RSNA PE was used as the derivation and internal validation data for PE classification models. The CTPA dataset was used for validating the transferability of PE models' in predicting Chronic thromboembolic pulmonary disease (CTEPD) in 3-6 months.}
\label{tab:cohorts}
\begin{tabularx}{\linewidth}{lCCC}
\toprule
\textbf{Characteristic} & \textbf{COPD-CT-DF} & \textbf{RSNA PE} & \textbf{CTPA} \\
\midrule
\multicolumn{4}{c}{\textit{Data source and acquisition}} \\ \hline
Source & Single-center & Multi-center (5 countries) & Single-center \\
Disease / classification task & COPD diagnosis & Pulmonary embolism diagnosis & Pulmonary embolism outcome \\
Imaging protocol & Standard chest CT & CT pulmonary angiography & CT pulmonary angiography \\
Slice thickness (mm) & 1.0--1.25 & 2.5--3.0 & 1.0--1.25 \\
\midrule
\multicolumn{4}{c}{\textit{Cohort composition}} \\ \hline
Total cases (n) & 719 & 1{,}433 (random subset) & 161 \\
Training / Validation / Test (n) & 526 / 93 / 100 & 712 / 356 / 365 & 129 / 32 / -- \\
Positive cases, n (\%) & 365 (50.8) & 533 (37.2) & 118 (73.3) \\
Negative cases, n (\%) & 354 (49.2) & 900 (62.8) & 43 (26.7) \\
\bottomrule
\end{tabularx}
\end{subtable}

\vspace{1.2em}

\begin{subtable}[t]{\linewidth}
\centering
\caption{Clinical characteristics of the COPD-CT-DF cohort, stratified by COPD diagnosis.}
\label{tab:copd_clinical}
\begin{tabularx}{\linewidth}{lCC}
\toprule
\textbf{Characteristic} & \textbf{COPD (n=365)} & \textbf{Non-COPD (n=354)} \\
\midrule
\multicolumn{3}{c}{\textit{Demographics}} \\ \hline
Age (years) & 71.3 $\pm$ 10.2 & 66.5 $\pm$ 12.8 \\
Male, n (\%) & 273 (74.8) & 186 (52.5) \\
BMI (kg/m$^2$) & 23.1 $\pm$ 3.6 & 24.3 $\pm$ 6.9 \\
\midrule
\multicolumn{3}{c}{\textit{Smoking history, n (\%)}} \\ \hline
\quad Current smoker & 158 (43.3) & 89 (25.1) \\
\quad Former smoker & 29 (7.9) & 38 (10.7) \\
\quad Never smoker & 100 (27.4) & 111 (31.4) \\
\quad Unknown & 78 (21.4) & 116 (32.8) \\
\midrule
\multicolumn{3}{l}{\textit{Pulmonary function}} \\
FEV$_1$ (\% predicted) ($P<$0.001) & 60.8 $\pm$ 19.4 & 91.2 $\pm$ 21.8 \\
FEV$_1$/FVC (\%) ($P<$0.001) & 60.3 $\pm$ 8.1 & 82.1 $\pm$ 7.3 \\
\midrule
\multicolumn{3}{c}{\textit{COPD severity (GOLD), n (\%)}} \\ \hline
\quad GOLD 1 (Mild) & 44 (12.1) & -- \\
\quad GOLD 2 (Moderate) & 211 (57.8) & -- \\
\quad GOLD 3 (Severe) & 91 (24.9) & -- \\
\quad GOLD 4 (Very Severe) & 19 (5.2) & -- \\
\bottomrule
\end{tabularx}
\end{subtable}

\end{table*}

This study repositions knowledge distillation as a mechanism for cross-window feature integration and anatomical reasoning. Unlike input-level fusion (e.g., multi-channel paradigms), we train independent models for each window to preserve window-specific feature extraction pathways. We then align these models through distillation in deep feature space (2048 dimensions), enabling student models to ``learn'' latent pathological patterns that are not salient in their own input (effectively invisible) but clear in the teacher's domain/window. The distillation loss forces student features to approximate teacher features, representing not simple output mimicry but a deep representation alignment in abstract semantic space, enabling the student to identify structural proxies for systemic disease that traditional single-window or late-fusion approaches fail to capture.

\section{Materials and Methods}

This retrospective study used three datasets for two pulmonary disease diagnosis tasks of COPD and Pulmonary Embolism, and one prediction task of Chronic thromboembolic pulmonary disease (CTEPD) (see Table~\ref{tab:dataset_characteristics}). The COPD-CT-DF and CTPA datasets were collected at a tertiary academic medical centre under ethical approval granted by the institutional ethics committee (approval date: 2025), with data collected between March 2018 and December 2020. All patient data were de-identified prior to analysis, and the study was conducted in accordance with the Declaration of Helsinki. The RSNA PE dataset is a publicly available dataset and was used in accordance with its original release terms. 

For COPD diagnosis, the COPD-CT-DF dataset was used for the model derivation and internal validation. For PE diagnosis, a subset of RSNA PE was used for the derivation and internal validation. The CTPA dataset was used for validating the transferability of PE models in predicting CTEPD diseases in 3-6 months.

\subsection{Datasets}

\subsubsection{COPD-CT-DF Dataset}
The COPD-CT-DF dataset was drawn from the COPD-365 project at a tertiary academic medical centre, initially comprising 1,004 subjects who underwent both chest CT examinations and pulmonary function testing. Inclusion criteria were age $\geq$18 years, completion of a standard chest CT scan, complete pulmonary function test data (FEV$_1$ and FVC), and CT image quality meeting diagnostic standards. Exclusion criteria covered incomplete or non-axial CT scans, concomitant serious lung diseases (lung cancer, interstitial lung disease, active tuberculosis, bronchiectasis), missing key clinical data, and history of chest surgery or major thoracic deformities. After systematic quality control and screening, the final cohort consisted of 719 patients (inclusion rate: 71.6\%), comprising 365 COPD patients and 354 non-COPD controls.

COPD was diagnosed according to the Global Initiative for Chronic Obstructive Lung Disease (GOLD) criteria \citep{agusti2023global}, defined as a post-bronchodilator FEV$_1$/FVC ratio $<$0.70. Disease severity was stratified according to GOLD grades 1--4 based on FEV$_1$\% predicted. The control group consisted of subjects with FEV$_1$/FVC $\geq$0.70 and no history of chronic respiratory diseases. The two groups differed significantly in age, sex distribution, and pulmonary function parameters ; detailed cohort characteristics are reported in the bottom sub-figure of Table~\ref{tab:dataset_characteristics}. All chest CT examinations were performed using a standardized protocol: tube voltage 120 kVp, automatic tube current modulation, slice thickness 1.0-1.25 mm, and reconstruction interval 0.625-1.0 mm, with images acquired in the supine position at full inspiration and stored in DICOM format.

\subsubsection{RSNA PE Dataset}
The RSNA PE dataset is the publicly available RSNA Pulmonary Embolism CT dataset \citep{colak2021rsna}, a widely used public benchmark for pulmonary embolism detection, comprising chest CT pulmonary angiography images and corresponding PE annotations aggregated from multiple international medical centers. The original dataset contains substantially more negative than positive cases; to match the computational resources and training protocol of this study while alleviating this class imbalance, we performed patient-level undersampling of the negative cases, yielding a working subset of 1{,}433 patients with a positive-to-negative ratio of 36.7\% / 63.3\%. This subset was further split at the patient level into training (n=712), validation (n=356), and held-out test (n=365) sets. Both the sampling and the split used a fixed random seed (seed=42) to ensure reproducibility and prevent data leakage. CT images were used according to the original release settings, without additional resampling or protocol harmonization.

\subsubsection{CTPA Dataset}
The CTPA dataset was collected at a tertiary academic medical centre and was used to assess the cross-task transferability of the proposed framework on a related clinical task. Unlike the acute PE detection task targeted by the RSNA PE dataset, this dataset focuses on the clinical outcome adjudication of patients with a confirmed PE diagnosis. Each patient underwent a baseline CTPA scan at the time of acute PE diagnosis and a follow-up CTPA scan 3--6 months later. Clinicians classified each patient into one of two categories based on the imaging findings of the follow-up CTPA: resolved acute PE (complete or substantial thrombus dissolution with restored blood flow) or progression to chronic thromboembolic pulmonary disease (CTEPD; characterized by organized thrombi, vascular web formation, or vascular stenosis). The model input consists of the baseline CTPA scan, and the task is to predict the patient's final clinical outcome based solely on the initial imaging. This setup simulates the clinical need to assess the risk of thrombus evolution from a single CT examination. Given the limited sample size (161 patients; training set n=129, validation set n=32), the validation set was used as the final evaluation set, on which the transfer learning and fine-tuning results were reported. All CT images were acquired using routine clinical CTPA protocols and stored in DICOM format.

\subsection{Image Preprocessing}
\label{sec:Image Preprocessing}
The three datasets shared a unified preprocessing pipeline, differing only in the number of sampled slices to accommodate task-specific characteristics. For each patient, DICOM files were loaded and CT slices were standardized to consistent anatomical directions based on the ImagePositionPatient metadata, ensuring proper cranio-caudal orientation from apex to base. Raw pixel values were converted to Hounsfield Units (HU) using the RescaleSlope and RescaleIntercept parameters from DICOM headers. To exclude incomplete anatomical coverage at scan boundaries, the first and last 10\% of slices were removed from each patient's CT series.

The slice sampling strategy was adapted to the nature of each task. For the COPD-CT-DF dataset, given that COPD is a diffuse disease whose pathological changes are relatively uniformly distributed throughout the lung parenchyma, we defined a region starting at 40\% from the apex in the cranio-caudal direction to emphasize the central lung regions where COPD-related changes are most prominent \citep{jeffery1998structural}, and performed unbiased sampling of exactly 32 representative slices within this region. For the RSNA PE and CTPA datasets, because pulmonary embolism is a focal disease in which emboli may occur in any branch of the pulmonary arterial tree, we performed unbiased sampling of 128 slices across the entire lung to ensure complete coverage of the pulmonary vasculature and to avoid missing critical lesions. For patients with an insufficient number of slices, the sampling range was expanded accordingly to reach the target slice count.

Each selected slice was resized to 512$\times$512 pixels using bilinear interpolation while preserving anatomical proportions. To enhance the visibility of different anatomical structures and pathological features, five task-specific CT window settings were applied to each slice. Four window settings were shared across both tasks: (1) \textbf{Lung window} (width: 1500 HU, level: $-$600 HU) optimized for visualizing pulmonary parenchyma, emphysema, and air-filled structures; (2) \textbf{Mediastinal window} (width: 350 HU, level: 20 HU) designed for soft tissues, blood vessels, and mediastinal structures; (3) \textbf{High-resolution CT (HRCT) window} (width: 2000 HU, level: $-$600 HU) providing enhanced detail for fine lung parenchymal structures and subtle abnormalities; (4) \textbf{Zero window} (width: 1500 HU, level: 0 HU) serving as an intermediate setting for comprehensive tissue visualization. The fifth window was task-specific: for the COPD-CT-DF dataset, we used a \textbf{Bone window} (width: 1000 HU, level: 250 HU) tailored for osseous structures including ribs and vertebrae; for the RSNA PE and CTPA datasets, we used a \textbf{PE window} (width: 700 HU, level: 100 HU) specifically designed to enhance the contrast between intravascular filling defects and opacified pulmonary arteries for embolism detection~\citep{huhtanen2022automated}. Each window transformation was applied using the standard formula: $I_{\text{windowed}} = \frac{\text{clip}(I_{\text{raw}}, L-W/2, L+W/2) - (L-W/2)}{W}$, where $I_{\text{raw}}$ represents the raw CT intensity in Hounsfield Units, $W$ denotes window width, and $L$ denotes window level.

Finally, all windowed images were normalized using z-score normalization based on the global statistics of the corresponding training set: $I_{\text{normalized}} = \frac{I_{\text{windowed}} - \mu}{\sigma + \epsilon}$, where $\mu$ and $\sigma$ denote the mean and standard deviation of the respective training set, and $\epsilon = 10^{-8}$ prevents division by zero. The complete preprocessing pipeline transforms each patient's raw DICOM series into a standardized 3D volume of shape $(1, T, 512, 512)$ for each of the five window settings, where $T=32$ for the COPD-CT-DF dataset and $T=128$ for the RSNA PE and CTPA datasets.

\begin{figure*}[t]
\centering
\includegraphics[width=\textwidth]{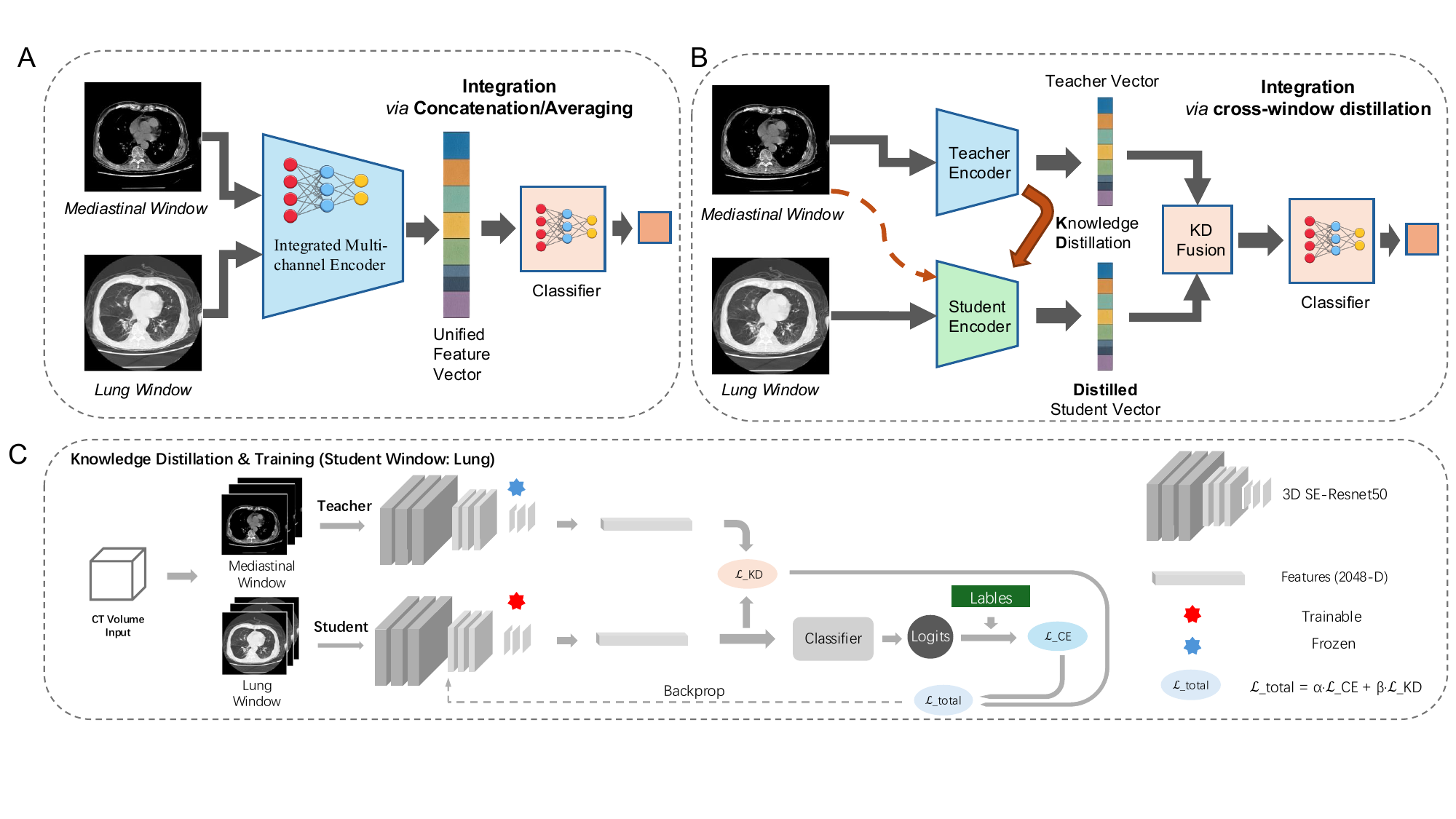}
\caption{\textbf{Conceptual Comparison and Technical Architecture of the Cross-Window Knowledge Distillation Framework.} (A) \textit{Baseline Integration via Concatenation/Averaging:} Standard multi-channel paradigms typically feed disparate viewing windows (e.g., Mediastinal and Lung) into a single unified encoder. This often results in feature interference and signal averaging within the Unified Feature Vector, potentially leading to diagnostic ambiguity. (B) \textit{Proposed Integration via Cross-Window Distillation:} Our framework decouples feature expertise by employing separate encoders for the Teacher (Mediastinal) and Student (Lung) domains. \textbf{The dashed arrow indicates the indirect Knowledge Distillation path. Rather than feeding Mediastinal pixels into the Student, this mechanism transfers high-level clinical understanding and structural priors from the Teacher.} This allows the Distilled Student Vector to identify latent features in the lung parenchyma that correlate with systemic pathology. (C) \textit{Knowledge Distillation and Training Pipeline:} A technical overview of the learning process for a specific student window (e.g., Lung). Architecture: Both pathways utilize a 3D SE-ResNet50 backbone to process 2048-D feature vectors. Training Dynamics: The Teacher model is pretrained and frozen (blue star), providing stable expert guidance, while the Student model remains trainable (red star). Loss Optimization: The framework is optimized via Backprop using a total loss function ($\mathcal{L}_{total}$). This function balances the standard cross-entropy loss ($\mathcal{L}_{CE}$) from ground-truth labels with the knowledge distillation loss ($\mathcal{L}_{KD}$), which minimizes the discrepancy between student and teacher logits.
}
\label{fig:architecture}
\end{figure*}

\subsection{Model Architecture}
Unlike standard multi-window paradigms that rely on input-level fusion (Figure~\ref{fig:architecture}A), our architecture decouples anatomical expertise to prevent feature interference and signal dilution. In conventional approaches, disparate voxel densities from different viewing windows are typically concatenated or averaged into a single multi-channel encoder. This ``shallow'' integration forces the network to process conflicting textural information at the earliest stages, often missing subtle, window-specific pathological nuances. In contrast, our proposed framework (Figure~\ref{fig:architecture}B) employs a dual-encoder strategy where teacher and student models independently develop high-level semantic representations for their respective windows. By utilising a cross-window knowledge distillation path (indicated by the dashed arrow in Figure~\ref{fig:architecture}B), the student model is guided by the teacher’s clinical priors rather than its raw pixel data. This enables the student to internalise latent pathological signatures—features effectively invisible in its own input setting—by aligning its feature representation with the teacher’s expert understanding in a shared abstract semantic space.

Technically, we employed SE-ResNet50, a 3D convolutional neural network architecture from the MONAI framework (Cardoso MJ et al. arXiv preprint arXiv:2211.02701. 2022), as the shared backbone for all classification tasks in this study (Figure~\ref{fig:architecture}B--C). SE-ResNet50 integrates Squeeze-and-Excitation (SE) blocks into the standard ResNet50 architecture, enabling the model to recalibrate channel-wise feature responses through explicit modeling of channel interdependencies \citep{he2016identity, hu2018squeeze}. The network accepts 3D CT volumes with input shape $(1, T, 512, 512)$, where $T$ denotes the number of sampled slices, and outputs a 2048-dimensional feature vector from its final convolutional stage. The final classification layer was modified to produce a single output neuron with sigmoid activation for the corresponding binary classification task.

\subsection{Cross-Window Knowledge Distillation}
\label{sec:kd}

The goal of the proposed cross-window knowledge distillation framework is to integrate the complementary pathological information carried by different CT window settings into a unified feature space. Let $\mathcal{W} = \{w_1, \ldots, w_K\}$ denote the set of $K=5$ CT window transformations adopted in this study (Section~\ref{sec:Image Preprocessing}). For any patient $i$, the preprocessed DICOM series yields, under each window $w \in \mathcal{W}$, a three-dimensional volume $x_i^{(w)}$. We instantiate one SE-ResNet50 backbone $f_{\theta^{(w)}}(\cdot)$ per window, mapping an input volume to a $d=2048$ dimensional deep feature vector $h_i^{(w)} = f_{\theta^{(w)}}(x_i^{(w)})$, followed by a linear classification head $g_{\phi^{(w)}}(\cdot)$ producing the logit $z_i^{(w)}$. All single-window models are first trained independently under standard supervision with the binary cross-entropy loss. This stage yields $K$ supervised single-window models that together form the starting point of the subsequent distillation procedure.

Building on these $K$ models, we determine the teacher window in a data-driven manner. Unlike conventional knowledge distillation, which transfers knowledge from a high-capacity teacher to a low-capacity student, the teacher and student in our framework share an identical backbone architecture and differ only in their input window. Concretely, on each dataset, the teacher window $w^*$ is defined as the window that achieves the best classification performance on the validation set:
\[
w^{*} \;=\; \arg\max_{w \in \mathcal{W}} \; \text{AUC}_{\text{val}}\!\left(f_{\theta^{(w)}}, g_{\phi^{(w)}}\right),
\]
and the remaining $K-1$ windows $w_s \in \mathcal{W} \setminus \{w^*\}$ serve as student windows in the distillation stage. This criterion ensures that the teacher consistently provides the most reliable supervisory signal for downstream feature alignment; at the same time, because the density-space distribution of pathological phenotypes differs across datasets, $w^*$ is allowed to adapt to each dataset without any manual specification.

Once the teacher window is identified, distillation proceeds within a dual-pathway architecture (Figure~\ref{fig:architecture}B--C). The teacher network $f_{\theta^{(w^*)}}$ is inherited from the supervised pretraining stage and its parameters are fully frozen during student training, so as to provide a stable and deterministic supervisory signal; each student pathway corresponds to one student window $w_s$ to be trained. For each training sample $(x_i, y_i)$, the teacher and student pathways respectively receive two volumes $x_i^{(w^*)}$ and $x_i^{(w_s)}$ of the same anatomical region of the same patient under different window transformations, and forward-propagate them into their $2048$-dimensional feature vectors $h_i^{(w^*)}$ and $h_i^{(w_s)}$. The student parameters $\theta^{(w_s)}$ and its classification head $\phi^{(w_s)}$ are then jointly optimized under two complementary losses.

Specifically, the first term preserves the student's own discriminative capability on the target task,
\[
\mathcal{L}_{\text{cls}} \;=\; \text{BCE}\!\left(\sigma(z_i^{(w_s)}), \, y_i\right), \qquad z_i^{(w_s)} = g_{\phi^{(w_s)}}\!\left(h_i^{(w_s)}\right),
\]
while the second term treats the teacher feature as a target and forces the student feature to approximate it in the $2048$-dimensional space,
\[
\mathcal{L}_{\text{KD}} \;=\; \frac{1}{d}\,\big\| h_i^{(w_s)} - h_i^{(w^*)} \big\|_2^2.
\]
The total loss for the student is a convex combination of the two,
\[
\mathcal{L}_{\text{total}} \;=\; \alpha\,\mathcal{L}_{\text{cls}} + \beta\,\mathcal{L}_{\text{KD}}, \qquad \alpha = \beta = 0.5,
\]
where the weighting coefficients were determined through a preliminary search on the validation set. Crucially, $\mathcal{L}_{\text{KD}}$ acts on the deep feature $h$ rather than on the output logit $z$, and this choice is what distinguishes the proposed framework from conventional soft-label distillation.

The reason for imposing feature alignment on the deep semantic space rather than on the output layer lies in the essential difference between what the two constraints enforce. When the distillation constraint is placed at the output layer, the student is only required to agree with the teacher at the decision boundary, which amounts to soft-label mimicry at the prediction level. When the constraint is instead placed at the feature layer, the student must reconstruct, from its own input $x_i^{(w_s)}$, an internal representation consistent with the teacher's feature $h_i^{(w^*)}$, even though the two inputs differ substantially at the pixel level. This mechanism drives the student network to learn pathological patterns that are latent or subtle in its own window but salient in the teacher window, thereby achieving knowledge transfer at the level of feature representations rather than at the level of prediction alignment. The proposed cross-window distillation thus offers a deep, trainable, and dataset-adaptive feature-level fusion mechanism for integrating complementary multi-window information.

\subsection{Ensemble via Meta-Learner}
\label{sec:ensemble}

To further integrate the predictions of multiple window-specific models, we adopt a meta-learner-based ensemble strategy \citep{mienye2022survey}. The ensemble follows a two-level architecture. At the first level, the $K$ base models are kept fixed and forward-propagated on the validation set, producing, for each sample $i$, a vector of window-level predicted probabilities $\mathbf{p}_i = [\, p_i^{(w_1)}, \ldots, p_i^{(w_K)}\,] \in [0,1]^{K}$, where $p_i^{(w)} = \sigma(z_i^{(w)})$. At the second level, these probabilities serve as input features to a logistic-regression meta-learner $g(\cdot)$ \citep{thomas2023deep}, whose output gives the final ensemble prediction $\hat{y}_i = g(\mathbf{p}_i)$. Compared with simple equal-weight voting or averaging, this meta-learner adaptively assigns weights according to the empirical performance of each window on the validation set and learns non-trivial interactions between their predictions \citep{wu2023bayesian}. The meta-learner is trained solely on the validation set to avoid data leakage, and at test time both the base models and the meta-learner are frozen.

To evaluate the contribution of cross-window distillation to ensemble performance, we construct two comparable pipelines under the same meta-learner architecture: a \emph{supervised ensemble}, which uses $K$ independently trained single-window supervised models as base learners, and a \emph{distilled ensemble}, which instead uses the teacher model and the $K-1$ distilled student models from Section~\ref{sec:kd}. The two pipelines differ only in whether the base models have been trained with cross-window distillation; all other structural and protocol choices are kept identical, so that any performance gap can be attributed directly to the feature-level distillation itself.

\subsection{Studying Cross-Window Information Transfer Mechanisms}
A central hypothesis of this study is that disparate clinical features from multiple CT window settings can be more effectively integrated through novel knowledge-distillation architectures than through standard concatenation. To evaluate the mechanism of cross-window information gain, we compared our Knowledge Distillation (KD) Fusion strategy against a standard Multi-channel Fusion baseline. 

\noindent \textbf{Validation Cohort: }We curated a domain-expert annotated dataset of 30 participants from the COPD-365 cohort. Two senior respiratory clinicians independently identified COPD-related abnormalities (e.g., vascular remodeling, airway thickening) across two separate viewing settings: the Mediastinal window and the Lung window. We specifically investigated how high-level radiological findings primarily visible in the Mediastinal window—the `Teacher' domain—inform the final diagnostic reasoning. 

Patients were stratified by the presence of these mediastinal findings, and two experimental comparisons were implemented:

\begin{itemize}
    \item \textbf{Full-System Fusion (Integration Logic):} We compared the KD Fusion system against a standard Multi-channel baseline, with both models having access to all imaging windows. This validates the system's ability to resolve feature conflicts and integrate complementary signals without the `averaging' effect common in multi-channel inputs.
    \item \textbf{Single-Window Ablation (Mechanism of Transfer):} To isolate the transfer of expertise, we compared the Distilled Student against a Standard Supervised baseline, restricting both models to Lung-window inputs only. This evaluates the internalisation of latent features, testing if the student can identify structural signatures that are typically `invisible' in the lung parenchyma without guidance from the Mediastinal Teacher.
\end{itemize}

Detailed descriptions of the validation cohort derivation, radiological features and annotation protocol are available at Appendix 1.

\subsection{Experimental Setup and Evaluation}

All models were implemented using PyTorch and trained on NVIDIA 4090 GPUs with batch size of 4. The Adam optimizer was employed with an initial learning rate of 0.001 and cosine annealing schedule over 40 epochs. Early stopping with patience of 10 epochs was applied based on validation loss to prevent overfitting. Binary cross-entropy loss was used for baseline model training, while knowledge distillation employed the combined loss function described previously. Data augmentation was not applied to maintain consistency with clinical CT acquisition protocols.

Model performance was evaluated using four standard metrics: accuracy, F1-score, recall, and area under the receiver operating characteristic curve (AUC). All metrics were computed on the held-out test set with 95\% confidence intervals calculated using bootstrap resampling (1000 iterations). Statistical comparisons between models were performed using paired t-tests, with $P < 0.05$ considered statistically significant \citep{jane2024guide}. The primary evaluation focused on AUC as it provides a comprehensive assessment of classifier performance across all decision thresholds, which is particularly relevant for clinical decision-making where sensitivity-specificity trade-offs must be considered.

\section{Results}

\subsection{Baseline Performance of Supervised Learning}
\label{sec:baseline}

Under the supervised setting without knowledge distillation, all $K=5$ single-window models were trained independently on each dataset, and the teacher window was determined for each dataset according to the validation-AUC criterion defined in Section~\ref{sec:kd}. The complete results are summarized in Table~\ref{tab:combined_distillation_comparison}.

On the \textbf{COPD-CT-DF} dataset, the five single-window supervised models exhibited substantial differences in discriminative performance. The mediastinal window yielded the best results across all evaluation metrics, with an accuracy of 0.8300 (95\% CI: 0.7501, 0.9032) and an AUC of 0.8960 (95\% CI: 0.8274, 0.9524), and was therefore designated as the teacher window for this dataset. The AUCs of the remaining four student windows ranged from 0.7467 to 0.8111, all below the teacher baseline. Among them, the bone window achieved the highest student baseline (AUC 0.8111; 95\% CI: 0.7219, 0.8880), followed by the lung window (AUC 0.7835; 95\% CI: 0.6895, 0.8655), with the HRCT and zero windows yielding AUCs of 0.7739 and 0.7467, respectively.

On the \textbf{RSNA PE} dataset, the PE window achieved the best overall performance among all five windows, with an accuracy of 0.8099 (95\% CI: 0.7713, 0.8512) and an AUC of 0.8819 (95\% CI: 0.8449, 0.9174), and was accordingly designated as the teacher window. The AUCs of the remaining four student windows ranged from 0.7952 to 0.8310. Among them, the zero window attained the highest student baseline (AUC 0.8310; 95\% CI: 0.7880, 0.8726), followed by the mediastinal window (AUC 0.8173; 95\% CI: 0.7690, 0.8605), the HRCT window (AUC 0.7953; 95\% CI: 0.7435, 0.8406), and the lung window (AUC 0.7952; 95\% CI: 0.7419, 0.8388). The AUC gap between each student window and the teacher window remained below 0.09, and the overall inter-window spread was narrower than that observed on the COPD-CT-DF dataset.

\subsection{Knowledge Distillation Improves Diagnostic Performance}

Cross-window knowledge distillation significantly improved diagnostic performance across all student windows on both datasets (Table~\ref{tab:combined_distillation_comparison}). On the \textbf{COPD-CT-DF} dataset, the HRCT window achieved the largest improvement, with AUC increasing from 0.7739 to 0.9384 (+0.1645; 95\% CI: 0.8896, 0.9752) and accuracy from 0.7000 to 0.8700 (+0.1700; 95\% CI: 0.8000, 0.9300); the lung, zero, and bone windows yielded AUC gains of +0.1561, +0.1541, and +0.1013, respectively. All four distilled student models ultimately achieved AUC values above 0.90, significantly outperforming their supervised baselines ($P<.001$). On the \textbf{RSNA PE} dataset, the distillation-induced AUC gains ranged from +0.0668 to +0.1141, with the HRCT window yielding the largest gain (+0.1141), followed by the lung window (+0.1003), the mediastinal window (+0.0996), and the zero window (+0.0668); the AUCs of the four distilled students ranged from 0.8955 to 0.9169, likewise significantly outperforming their supervised baselines ($P<.001$). The consistent improvement observed on both datasets indicates that the gain brought by cross-window distillation does not depend on a specific disease or a specific window configuration.

To further clarify the source of these improvements, we performed a sample-level agreement and attention analysis between the distilled and supervised models on the COPD-CT-DF test set. A Venn-diagram analysis showed that the distilled models corrected 21--24 cases per student window that had been misclassified by the supervised baselines, while maintaining a high degree of agreement with the supervised models (57--63 jointly correct predictions) and introducing only a small number of new errors (7--13 cases). Grad-CAM visualization \citep{selvaraju2017grad} further indicated that the distilled models developed more focused and disease-relevant attention patterns compared with their supervised counterparts. The complete mechanistic analysis and visualizations are reported in Appendix 2. Overall, by aligning heterogeneous window representations in the deep feature space, cross-window knowledge distillation successfully extracts complementary disease information from different CT windows and translates it into improved discriminative performance of the student models.

\begin{table*}[!htbp]
\centering
\caption{Comparison of Supervised Learning and Knowledge Distillation on COPD-CT-DF and RSNA PE Detection Datasets. For COPD-CT-DF, the teacher model is trained on mediastinal window (Acc=0.8300, F1=0.8172, Recall=0.7755, Precision=0.8636, AUC=0.8960). For RSNA PE, the teacher model is trained on PE window (Acc=0.8099, F1=0.7396, Recall=0.7153, AUC=0.8819). Student models with different CT windows are trained either by supervised learning alone or by knowledge distillation from the teacher. For COPD dataset, COPD is labeled as positive class (1). Values in parentheses indicate the improvement of distillation over supervised learning. Bold values indicate performance under distillation.}
\label{tab:combined_distillation_comparison}
\resizebox{\textwidth}{!}{%
\begin{tabular}{lcccccccccc}
\toprule
\multirow{2}{*}[0.55ex]{Window} & \multicolumn{2}{c}{Acc} & \multicolumn{2}{c}{F1} & \multicolumn{2}{c}{Recall} & \multicolumn{2}{c}{Precision} & \multicolumn{2}{c}{AUC} \\
\cmidrule(lr){2-3} \cmidrule(lr){4-5} \cmidrule(lr){6-7} \cmidrule(lr){8-9} \cmidrule(lr){10-11}
 & Supervise & Distillation & Supervise & Distillation & Supervise & Distillation & Supervise & Distillation & Supervise & Distillation \\
\midrule
\multicolumn{11}{l}{\textit{COPD-CT-DF dataset}} \\
Lung & 0.6700 & \textbf{0.8300} & 0.7130 & \textbf{0.8247} & 0.8367 & \textbf{0.8163} & 0.6212 & \textbf{0.8333} & 0.7835 & \textbf{0.9396} \\
 & & \textbf{(+0.1600)} & & \textbf{(+0.1117)} & & \textbf{($-$0.0204)} & & \textbf{(+0.2121)} & & \textbf{(+0.1561)} \\
\cmidrule{1-11}
Zero & 0.7000 & \textbf{0.7800} & 0.6154 & \textbf{0.7843} & 0.4898 & \textbf{0.8163} & 0.8276 & \textbf{0.7547} & 0.7467 & \textbf{0.9008} \\
 & & \textbf{(+0.0800)} & & \textbf{(+0.1689)} & & \textbf{(+0.3265)} & & \textbf{($-$0.0729)} & & \textbf{(+0.1541)} \\
\cmidrule{1-11}
HRCT & 0.7000 & \textbf{0.8700} & 0.6809 & \textbf{0.8632} & 0.6531 & \textbf{0.8367} & 0.7111 & \textbf{0.8913} & 0.7739 & \textbf{0.9384} \\
 & & \textbf{(+0.1700)} & & \textbf{(+0.1823)} & & \textbf{(+0.1836)} & & \textbf{(+0.1802)} & & \textbf{(+0.1645)} \\
\cmidrule{1-11}
Bone & 0.7100 & \textbf{0.8300} & 0.7071 & \textbf{0.8172} & 0.7143 & \textbf{0.7755} & 0.7000 & \textbf{0.8636} & 0.8111 & \textbf{0.9124} \\
 & & \textbf{(+0.1200)} & & \textbf{(+0.1101)} & & \textbf{(+0.0612)} & & \textbf{(+0.1636)} & & \textbf{(+0.1013)} \\
\midrule
\multicolumn{11}{l}{\textit{RSNA PE dataset}} \\
Lung & 0.7218 & \textbf{0.8375} & 0.5944 & \textbf{0.7900} & 0.5401 & \textbf{0.8102} & 0.6607 & \textbf{0.7708} & 0.7952 & \textbf{0.8955} \\
 & & \textbf{(+0.1157)} & & \textbf{(+0.1956)} & & \textbf{(+0.2701)} & & \textbf{(+0.1101)} & & \textbf{(+0.1003)} \\
\cmidrule{1-11}
Med & 0.7245 & \textbf{0.8430} & 0.6815 & \textbf{0.7865} & 0.7810 & \textbf{0.7664} & 0.6045 & \textbf{0.8077} & 0.8173 & \textbf{0.9169} \\
 & & \textbf{(+0.1185)} & & \textbf{(+0.1050)} & & \textbf{($-$0.0146)} & & \textbf{(+0.2032)} & & \textbf{(+0.0996)} \\
\cmidrule{1-11}
Zero & 0.7576 & \textbf{0.8320} & 0.6901 & \textbf{0.7782} & 0.7153 & \textbf{0.7810} & 0.6667 & \textbf{0.7754} & 0.8310 & \textbf{0.8978} \\
 & & \textbf{(+0.0744)} & & \textbf{(+0.0881)} & & \textbf{(+0.0657)} & & \textbf{(+0.1087)} & & \textbf{(+0.0668)} \\
\cmidrule{1-11}
HRCT & 0.7245 & \textbf{0.8375} & 0.6599 & \textbf{0.7839} & 0.7080 & \textbf{0.7810} & 0.6178 & \textbf{0.7868} & 0.7953 & \textbf{0.9094} \\
 & & \textbf{(+0.1130)} & & \textbf{(+0.1240)} & & \textbf{(+0.0730)} & & \textbf{(+0.1690)} & & \textbf{(+0.1141)} \\
\bottomrule
\end{tabular}
}
\end{table*}

\subsection{Model Ensemble Strategies}

Following the meta-learner-based ensemble framework described in Section~\ref{sec:ensemble}, we constructed two comparable pipelines for each dataset, a supervised ensemble and a distilled ensemble, and evaluated their final performance under the same meta-learner architecture (Table~\ref{tab:stacking_comparison}). On the \textbf{COPD-CT-DF} dataset, the distilled ensemble improved the accuracy from 0.7700 to 0.9400, the AUC from 0.8936 to 0.9960, the F1-score from 0.7647 to 0.9396, and the recall from 0.7673 to 0.9388, outperforming the supervised ensemble on every evaluation metric. On the \textbf{RSNA PE} dataset, the distilled ensemble likewise raised the accuracy from 0.7851 to 0.8733, the AUC from 0.8689 to 0.9382, the F1-score from 0.6486 to 0.8203, and the recall from 0.5255 to 0.7664, with the improvement in recall (+0.2409) being particularly pronounced, a gain of practical importance for pulmonary embolism detection, where missed diagnoses carry a high clinical cost. With the ensemble structure, the meta-learner, and the evaluation protocol held fixed, the performance gap between the two pipelines can be attributed directly to the feature-level distillation mechanism introduced in Section~\ref{sec:kd}.

\subsection{Cross-Task Transfer Learning Validation}

To further assess the transferability of the proposed framework, we transferred the ensemble pipelines trained on the RSNA PE dataset to the CTPA dataset and compared the supervised and the distilled ensembles under two settings (Fig~\ref{fig:transfer_comparison}): \emph{direct transfer}, in which the trained pipelines were applied to the CTPA validation set without any parameter update, and \emph{fine-tuned transfer}, in which the backbone networks and the meta-learner were kept frozen while only the classification heads of the base models were fine-tuned on the CTPA training set. Under direct transfer, the distilled ensemble yielded a higher AUC than the supervised ensemble (0.6603 vs.\ 0.5944). Under fine-tuned transfer, the advantage of the distilled ensemble became more pronounced, with the AUC increasing from 0.6264 to 0.7481 (+0.1217). Given the limited size of the CTPA validation set, these results should be interpreted as a preliminary observation of the framework's cross-task transferability; nevertheless, the consistent superiority of distillation-based pretraining over supervised pretraining under both transfer settings suggests that the deep feature representations learned through cross-window knowledge distillation generalize more effectively across related clinical tasks.

\subsection{Mechanistic Analysis of Information Transfer}
To investigate the origin of the observed performance gains, we conducted a sub-group analysis on 30 human-annotated cases. We categorised these cases by the presence of COPD-related regional, structural, and pathological markers predominantly observable in the Mediastinal window—the domain of our most predictive "\textit{Teacher}" model. Detailed lists of these markers, including vascular remodelling and right-heart strain indicators, are available in Appendix Tables A1–A3.

\begin{figure*}[htbp] 
     \centering
     \begin{subfigure}[t]{0.49\textwidth}
         \centering
         \includegraphics[width=\textwidth]{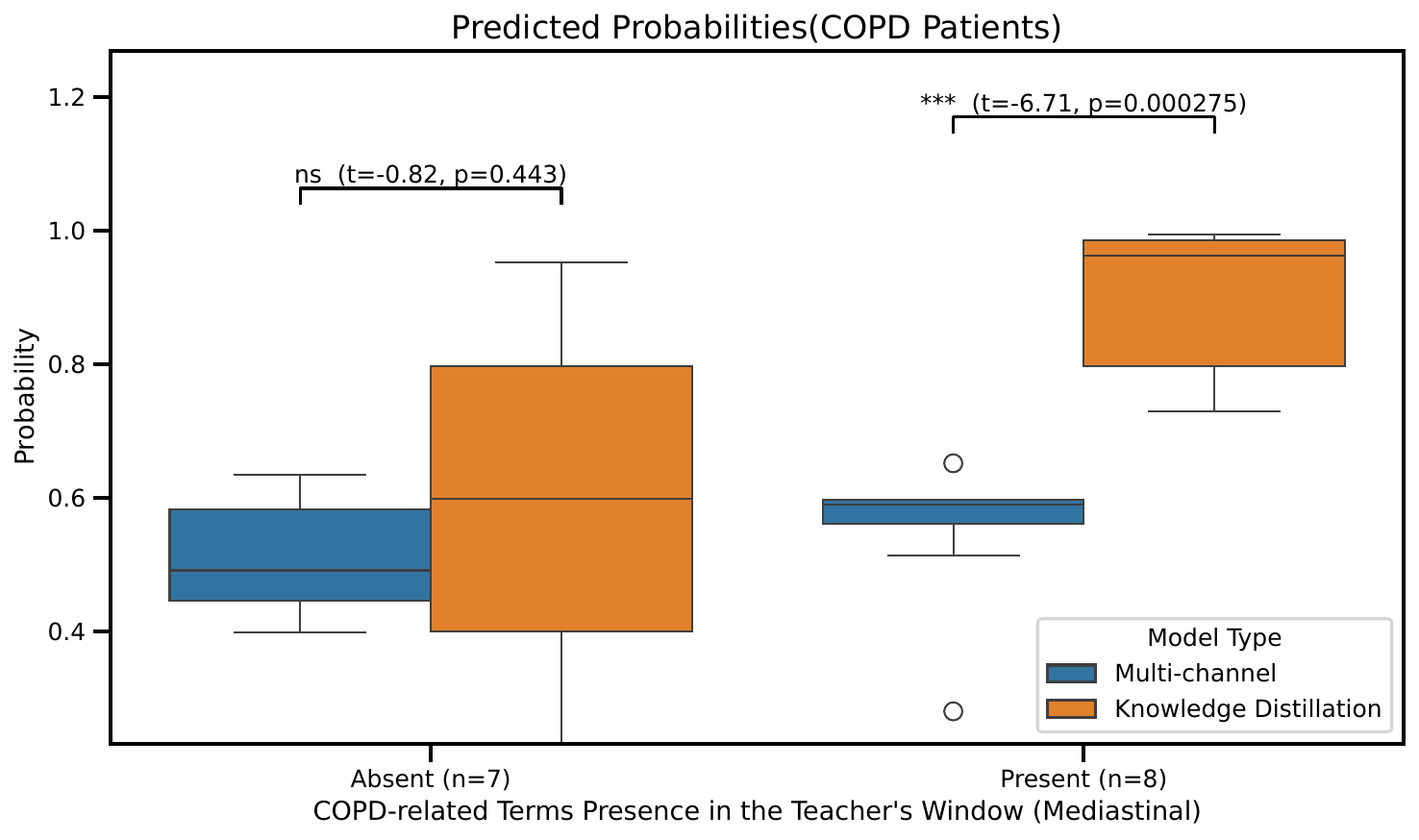}
         \caption{[Multi-channel vs. KD Fusion - COPD Patients]: Comparison of predicted probabilities for COPD patients using the full integrated models. Results are stratified by the presence of COPD-related structural markers in the Mediastinal (Teacher) window. Knowledge Distillation (KD) Fusion significantly resolves the ambiguity seen in the Multi-channel baseline ($p = 0.000275$) when mediastinal markers are present.}
         \label{fig:comp-fusion-copd}
     \end{subfigure}
     \hfill
     \begin{subfigure}[t]{0.49\textwidth}
         \centering
         \includegraphics[width=\textwidth]{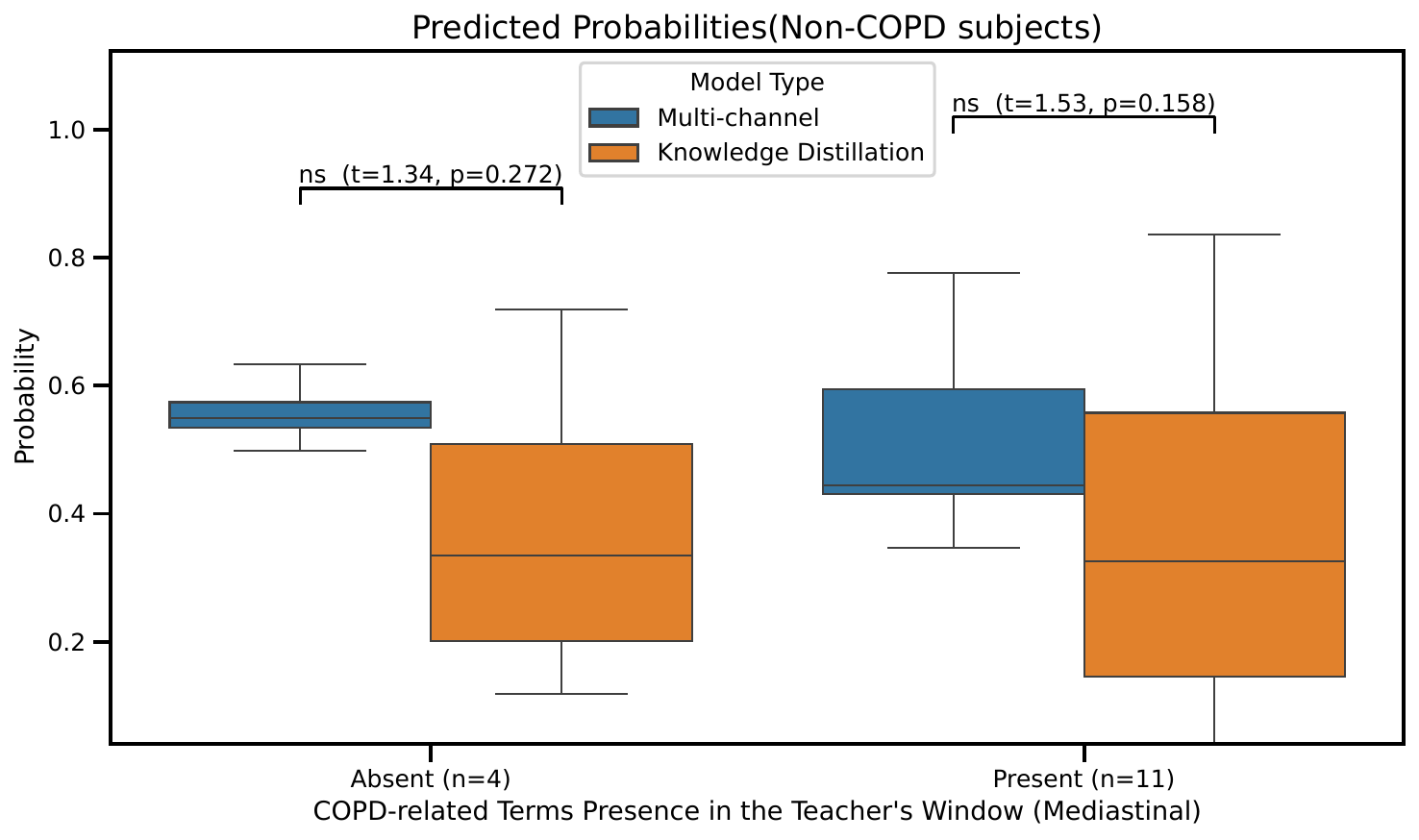}
         \caption{[Multi-channel vs. KD Fusion - Non-COPD Subjects]: Calibration of full integrated models on healthy subjects. The KD Fusion model exhibits superior specificity by suppressing false-positive tendencies observed in the Multi-channel baseline, maintaining lower median probabilities across both marker-absent and marker-present subgroups.}
         \label{fig:comp-fusion-non-copd}
     \end{subfigure}

     \vspace{0.2cm}
     \hdashrule{\textwidth}{0.5pt}{4pt 4pt} 
     \vspace{0.2cm}
     
     \begin{subfigure}[t]{0.49\textwidth}
         \centering
         \includegraphics[width=\textwidth]{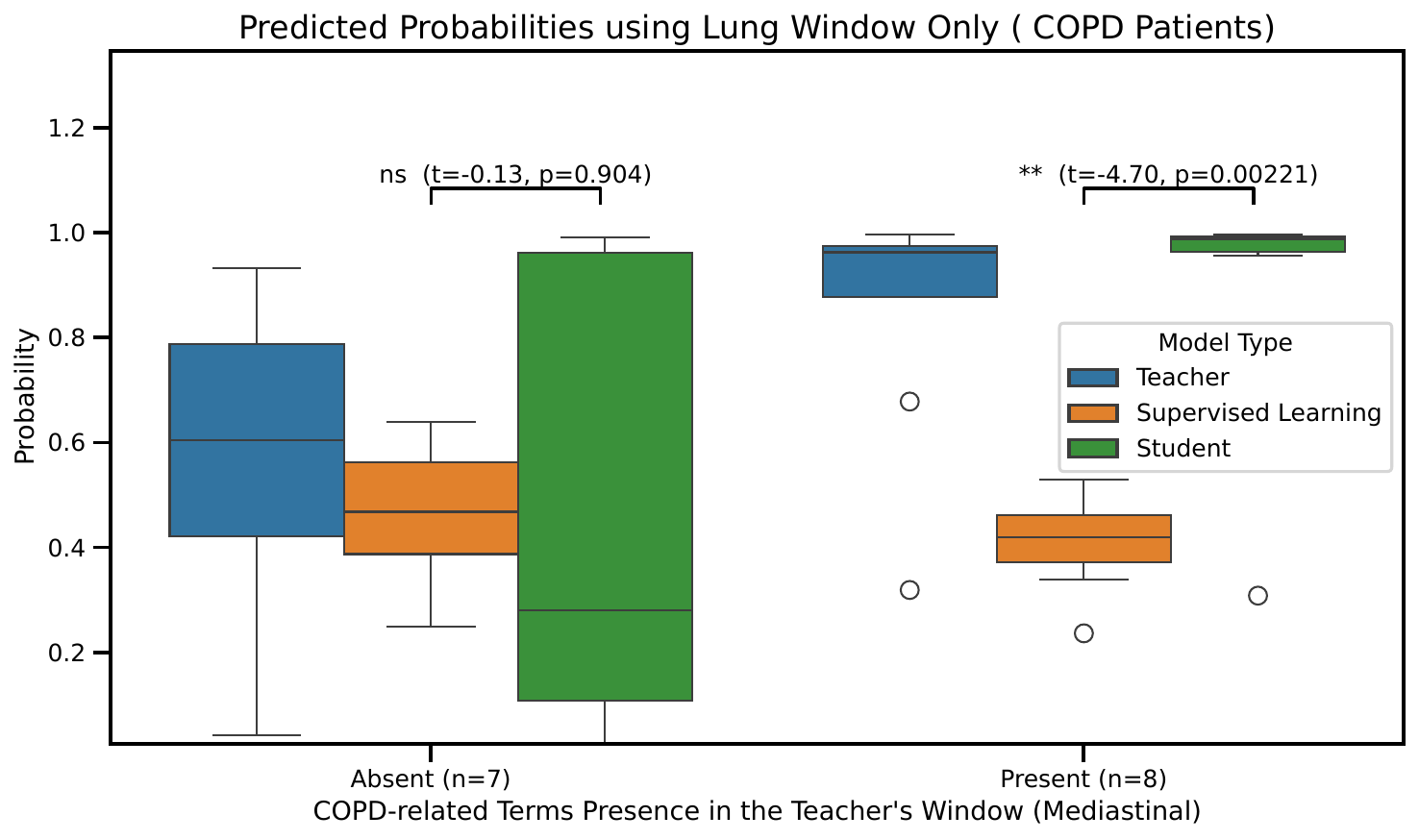}
         \caption{ [Lung Window Only - COPD Patients]: Ablation analysis comparing the Distilled Student and Standard Supervised models using only the lung window as input. In cases where disease markers are primarily visible in the mediastinal view, the Distilled Student identifies COPD with significantly higher sensitivity ($p = 0.00221$) than the supervised baseline, demonstrating successful transfer of latent diagnostic priors.}
         \label{fig:comp-lung-copd}
     \end{subfigure}
     \hfill
     \begin{subfigure}[t]{0.49\textwidth}
         \centering
         \includegraphics[width=\textwidth]{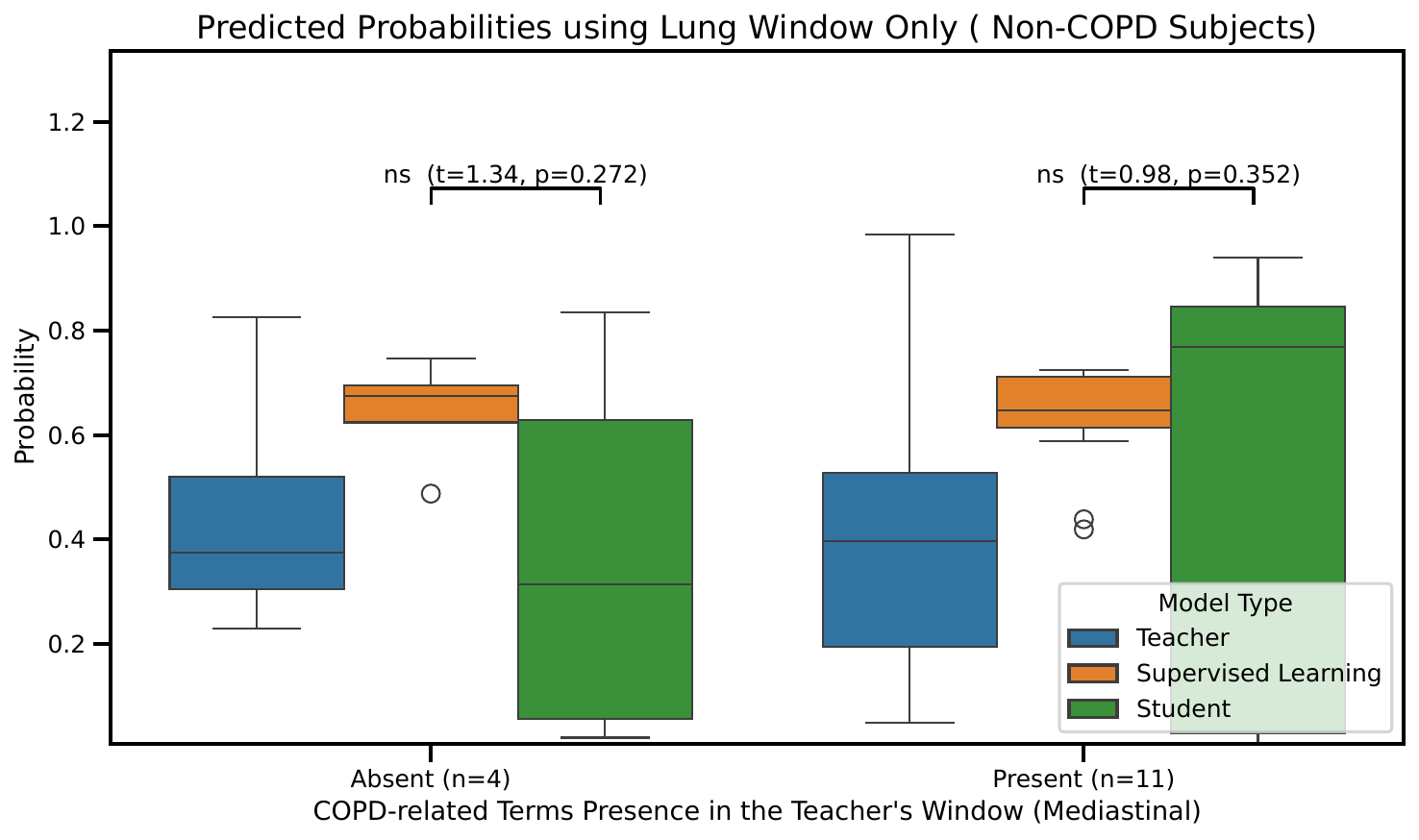}
         \caption{[Lung Window Only - Non-COPD Subjects]: Specificity validation of lung-view models on healthy subjects. The Distilled Student demonstrates higher skepticism toward parenchymal noise compared to the supervised baseline, indicating that teacher-guided training reduces incidental feature overfitting.}
         \label{fig:comp-lung-non-copd}
     \end{subfigure}

     \caption{Mechanistic Validation of Knowledge Distillation (KD) Fusion via Human-Annotated Phenotypes (n=30). (A-B) demonstrate the overall system performance, where the proposed KD Fusion architecture significantly outperforms a standard Multi-channel baseline by resolving feature interference in diseased patients and improving calibration in healthy subjects. (C-D) isolate the mechanism of this improvement through a lung-window-only ablation study. The results confirm that the Distilled Student internalizes "latent" structural expertise from the Mediastinal Teacher, allowing it to detect systemic disease signatures that are visually subtle or absent to a standard supervised model. This confirms that the performance gain is driven by an enriched latent feature space and superior anatomical reasoning rather than simple ensemble averaging.}
     \label{fig:comp_30_pts}
\end{figure*}

\begin{figure*}[!htbp]
  \centering
  \begin{subfigure}[t]{0.49\linewidth}
    \centering
    \includegraphics[width=\linewidth]{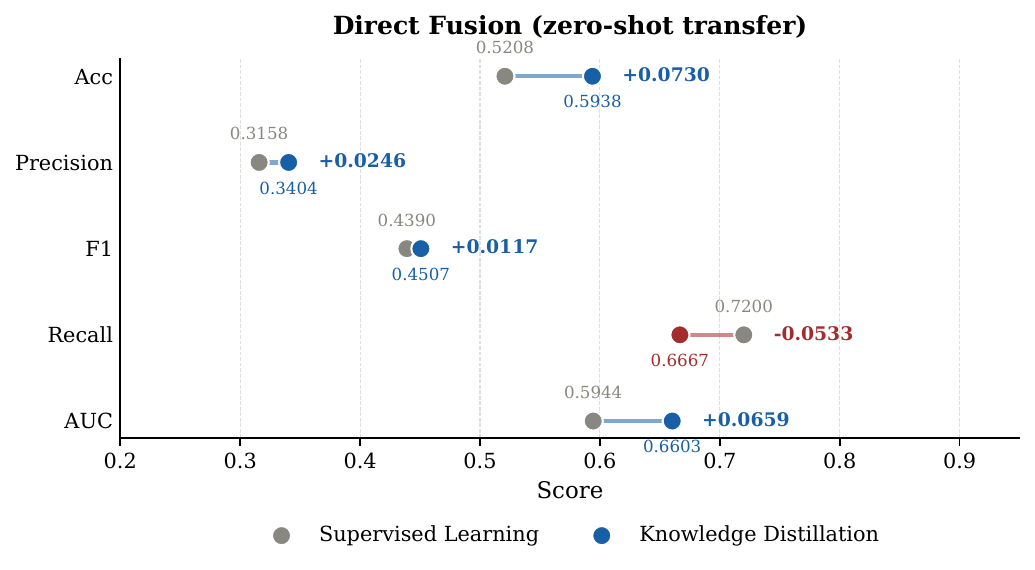}
    \caption{Direct transfer (zero-shot).}
    \label{fig:transfer_direct}
  \end{subfigure}
  \hfill
  \begin{subfigure}[t]{0.49\linewidth}
    \centering
    \includegraphics[width=\linewidth]{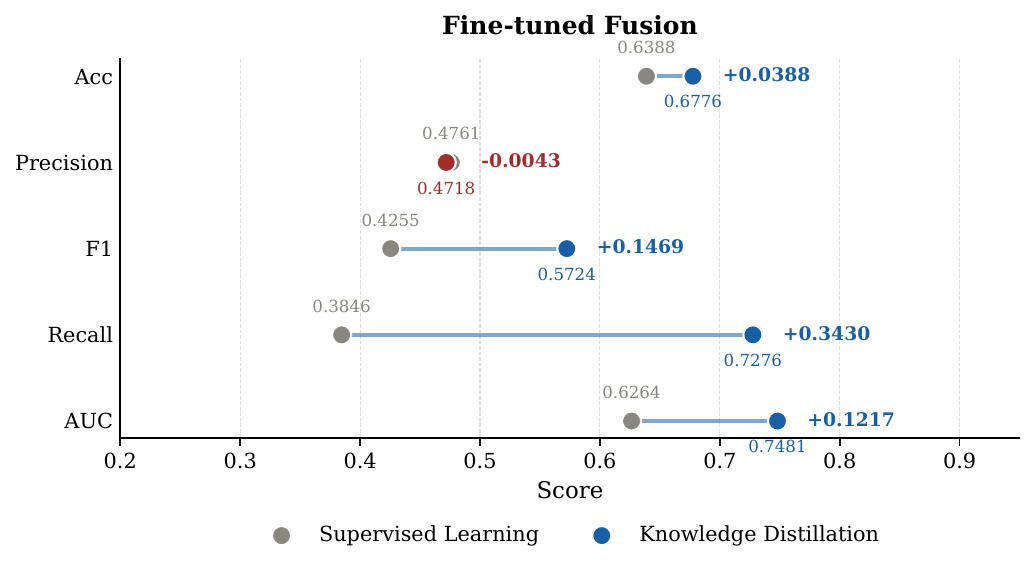}
    \caption{Fine-tuned transfer.}
    \label{fig:transfer_finetuned}
  \end{subfigure}
  \caption{Per-metric comparison of supervised learning and knowledge distillation when transferring from the RSNA PE dataset to the CTPA dataset, under (a) direct (zero-shot) transfer and (b) fine-tuned transfer. In each panel, gray markers denote the supervised baseline and blue markers denote the distilled model when KD outperforms SL; red markers indicate the rare cases in which KD underperforms SL. The annotated value to the right of each pair gives the absolute gain (KD $-$ SL).}
  \label{fig:transfer_comparison}
\end{figure*}

This analysis focuses on the model's ability to utilise these systemic markers to resolve diagnostic ambiguity and inform the final COPD classification (see Figure~\ref{fig:comp_30_pts}).

\noindent \textbf{Full-System Fusion (Integration Logic)} In the full-system comparison (Figures 2a–2b), our Knowledge Distillation (KD) Fusion architecture demonstrated superior integration of multi-window evidence compared to the standard Multi-channel baseline.
\begin{itemize}
\item \textbf{Enhanced Sensitivity:} For COPD patients with clear mediastinal markers, the Multi-channel baseline produced hesitant, near-threshold predictions (median $p \approx 0.58$). In contrast, the KD Fusion model resolved this ambiguity, achieving significantly higher diagnostic confidence (median $p \approx 0.95$, $p = 0.000275$).
\item \textbf{Calibration and Specificity:} In non-COPD subjects, the KD Fusion model showed improved calibration by maintaining lower probabilities (median $p \approx 0.31$) compared to the Multi-channel baseline (median $p \approx 0.55$). This suggests that the KD framework effectively filters incidental features that often lead to false-positive tendencies in standard multi-channel paradigms.
\end{itemize}

\noindent \textbf{Single-Window Ablation (Mechanism of Transfer)} To isolate the origin of these gains, we compared the Distilled Student against a Standard Supervised baseline using only lung-window inputs (Figures 2c–2d). This ablation reveals the "hidden" knowledge transfer enabled by the distillation process:
\begin{itemize}\item \textbf{Latent Feature Extraction:} In COPD cases where diagnostic evidence was localised to the mediastinal view, the standard Supervised model (lung-only) failed to detect the disease (median $p \approx 0.45$). However, the Distilled Student—processing the exact same lung-view pixels—correctly identified these cases with high confidence (median $p = 0.98$, $p = 0.00221$).
\item \textbf{Synthesis of Findings:} These results provide empirical evidence that the student has internalised a structural diagnostic prior. By recognising "proxies" within the lung parenchyma that correlate with systemic remodelling, the student identifies signatures that are visually latent to models trained via standard label-based supervision.
\end{itemize}

\section{Discussion}

\subsection{Principal Findings}
The central finding of this study is that the complementary information carried by different CT windows can be effectively integrated in the deep feature space rather than in the input space. With the backbone and training data held fixed, unidirectionally aligning each student-window feature to the teacher-window feature in a $2048$-dimensional semantic space yields significant and stable improvements on all student windows, independent of any specific disease or window configuration. This challenges the implicit assumption underlying existing multi-window methods that fusion is a data-representation problem, and suggests that inter-window complementarity need not be made explicit at the pixel level via multi-channel stacking, but can instead be implicitly reconstructed by the student network from its own input. Under a strict comparison in which the ensemble architecture and evaluation protocol are kept identical, the distilled ensemble still significantly outperforms its supervised counterpart, indicating that the observed gain must be attributed to the representational quality of the base models rather than to the aggregation mechanism. The cross-task transfer experiments further support this attribution: on a target task with limited samples, the representations obtained through distillation transferred more effectively than those from supervised pretraining, suggesting that what distillation learns is not an overfit to the source-task labels but a cross-window semantic structure that is relatively disease-independent.

\subsection{Methodological Innovation}
The methodological contribution of this study can be summarized as reformulating multi-window CT fusion as a problem of heterogeneous representation alignment. Traditional approaches oscillate between two paradigms: single-window preprocessing, which discards multi-window complementarity , and input-level multi-channel fusion, which forces disparate windows to be mixed before any window-specific representation has been established, producing shallow feature interference. Both share the limitation of treating windows as different observational channels of the same underlying signal. In contrast, the present work takes the view that different windows correspond to orthogonal dimensions of the disease phenotype space, and that their integration should therefore occur only after each window has developed its own feature-extraction pathway. Fusion is accordingly lifted into a $2048$-dimensional deep semantic space and enforced through a feature-matching loss, so that the student network learns to reconstruct, from its own window input, the representation carried by the teacher window under the same semantic geometry.

Complementing this reformulation is the data-driven selection of the teacher window. Existing cross-representation distillation studies typically designate the teacher a priori from domain knowledge, whereas in our framework the teacher is determined automatically from the validation performance of the single-window supervised models. This shifts the identity of the teacher from a disease-physiological decision to an empirically verifiable data decision: on the two datasets considered in this study, the optimal teacher windows differed, yet the resulting distilled students achieved consistent gains in both cases, empirically confirming the cross-dataset stability of the mechanism.

\subsection{Mechanistic Insights}

The primary driver of our model’s superior performance lies in its ability to overcome ``window-blindness'' through structured expertise transfer. Clinical interpretation traditionally separates the lung and mediastinal windows into parenchymal and cardiovascular domains, respectively. However, our mechanistic analysis of information transfer demonstrates that these domains are not mutually exclusive to a distilled model. By employing a Mediastinal ``\textit{Teacher}'' to guide a Lung ``\textit{Student}'', we effectively internalised a structural diagnostic prior within the lung-view encoder. Our single-window ablation study proves that the distilled student identifies COPD signatures that are visually latent to standard supervised models, even when processing identical pixels. This suggests that the distillation process forces the model to recognise subtle parenchymal shifts and vascular patterns that serve as reliable proxies for the systemic remodelling visible in the mediastinal view. Unlike standard multi-channel architectures, which often suffer from feature interference and produce ambiguous, averaged predictions ($p \approx 0.58$), our framework ensures that expert knowledge is preserved and injected, enabling the model to transition from simple pixel-pattern matching to integrated anatomical reasoning.

\subsection{Clinical Implications and Limitations}
The ability of the KD Fusion model to resolve diagnostic ambiguity across diverse pathologies has significant implications for the automated screening of both chronic and acute respiratory conditions. Our results on the COPD-CT-DF and RSNA PE datasets demonstrate that by leveraging cross-window knowledge, AI systems can achieve a level of sensitivity that closely mirrors multi-window expert review. In the context of COPD, the significant confidence gain ($p < 0.001$) in identifying systemic phenotypes—without sacrificing specificity—suggests a more reliable tool for longitudinal management.

Furthermore, the substantial performance gains observed in the RSNA PE dataset, particularly the marked improvement in recall (from $0.52$ to $0.76$), underscore the system's utility in acute diagnostic settings. Pulmonary embolism often presents with subtle vascular cues in the mediastinal window that may be overlooked when focus is restricted to parenchymal changes. By effectively transferring expertise between viewing planes, our framework minimises the risk of missed diagnoses in complex cases where pathological signatures are non-obvious or distributed across disparate anatomical regions. Ultimately, such a well-calibrated system reduces the potential for ``AI fatigue'' among radiologists by providing certain, clinically grounded outputs that align with the multi-dimensional nature of pulmonary vascular and airway diseases.

\begin{table*}[!htbp]
\centering
\caption{Performance comparison of Supervised Learning (SL) and Knowledge Distillation (KD) with Stacking fusion on the COPD-CT-DF and RSNA PE datasets. Bold values indicate KD performance; values in the right sub-column of each metric show the absolute improvement of KD over SL.}
\label{tab:stacking_comparison}
\resizebox{\textwidth}{!}{%
\begin{tabular}{ll cc cc cc cc cc}
\toprule
\multirow{2}{*}[0.55ex]{Strategy} & \multirow{2}{*}[0.55ex]{Dataset} & \multicolumn{2}{c}{AUC} & \multicolumn{2}{c}{Acc} & \multicolumn{2}{c}{Precision} & \multicolumn{2}{c}{F1} & \multicolumn{2}{c}{Recall} \\
\cmidrule(lr){3-4} \cmidrule(lr){5-6} \cmidrule(lr){7-8} \cmidrule(lr){9-10} \cmidrule(lr){11-12}
& & Value & $\Delta$ & Value & $\Delta$ & Value & $\Delta$ & Value & $\Delta$ & Value & $\Delta$ \\
\midrule
SL    & COPD-CT-DF & 0.8936 & --- & 0.7700 & --- & 0.7621 & --- & 0.7647 & --- & 0.7673 & --- \\
KD    & COPD-CT-DF & \textbf{0.9960} & \textbf{+0.1024} & \textbf{0.9400} & \textbf{+0.1700} & \textbf{0.9404} & \textbf{+0.1783} & \textbf{0.9396} & \textbf{+0.1749} & \textbf{0.9388} & \textbf{+0.1715} \\
\cmidrule{1-12}
SL    & RSNA PE    & 0.8689 & --- & 0.7851 & --- & 0.8470 & --- & 0.6486 & --- & 0.5255 & --- \\
KD    & RSNA PE    & \textbf{0.9382} & \textbf{+0.0693} & \textbf{0.8733} & \textbf{+0.0882} & \textbf{0.8824} & \textbf{+0.0354} & \textbf{0.8203} & \textbf{+0.1717} & \textbf{0.7664} & \textbf{+0.2409} \\
\bottomrule
\end{tabular}
}
\end{table*}

Major limitations include: (1) Single-center data (n=719) limits generalizability and requires external validation; (2) Fixed teacher strategy, where bidirectional distillation or dynamic teacher selection might further improve performance; (3) Evaluation of only binary classification, without GOLD staging or quantitative emphysema assessment; (4) Small test set (n=100) limits statistical power for subgroup analyses; (5) Grad-CAM provides post-hoc interpretability but does not guarantee feature alignment with true pathophysiology \citep{ennab2025advancing}.

\subsection{Future Directions}

Future research will focus on evolving this framework towards dynamic domain adaptation, where adaptive window width and level configurations can be learned according to characteristics of the clinical question and CT studies, rather than choosing those based on human eyes. To further validate the generalisability of this framework, we intedn to apply to a broader spectrum of thoracic pathologies, such as pulmonary embolism, lung cancer, and other pulmonary diseases. Additional, we plan to transit toward multimodal and multi-objective learning by integrating clinical metadata such as pulmonary function tests and smoking history and simultaneously quantify complex diseae patterns through intergrated anatomical reasoning.

In summary, cross-window knowledge distillation provides a novel feature-level fusion approach for multi-window medical image analysis, and this framework has the potential to extend to other multi-window or multimodal imaging analysis tasks.

\bibliographystyle{unsrt}
\bibliography{references}

\appendix

\setcounter{figure}{0}
\renewcommand{\thefigure}{A\arabic{figure}}

\setcounter{table}{0}
\renewcommand{\thetable}{A\arabic{table}}

\section{Validation Dataset and Annotation Protocol}
\label{appex:validation}
\subsection{Participant Selection and Sampling Strategy}

To evaluate the mechanistic performance of the proposed architecture, we curated a representative validation subset from the COPD-365 test cohort. We performed stratified random sampling of $N=30$ participants to ensure a balanced distribution across clinical and model-performance dimensions:
\begin{enumerate}
\item \textbf{Clinical Balance:} 50\% ($n=15$) were clinically diagnosed with COPD and 50\% ($n=15$) were non-COPD controls.
\item \textbf{Performance Stratification:} Participants were sampled across the model's probability spectrum, including high-confidence correct predictions, low-confidence ``borderline" cases, and clear misclassifications, ensuring the validation set challenged the model's integration logic.
\end{enumerate}

\subsection{Expert Annotation Protocol}
Two senior respiratory clinicians (with $>10$ years of experience) served as independent annotators. To eliminate bias, a rigorous double-blinding protocol was implemented:
\begin{itemize}
\item \textbf{Outcome Blinding:} Annotators were blind to the ground-truth COPD diagnosis and the AI model’s predicted probability for each patient.
\item \textbf{Window-Level Blinding:} Evaluation was conducted at the ``window level" rather than the ``patient level." Slices from High-Resolution CT (HRCT) Lung windows and Mediastinal windows were presented as independent tasks. Annotators were blinded to which windows belonged to which patient, preventing them from using cross-window context to infer findings.
\end{itemize}

\subsection{Feature Taxonomy for COPD Assessment}
Annotators evaluated the images based on a hierarchical taxonomy of features across three categories. These features were selected based on their established relevance to COPD diagnosis and their typical visibility in specific viewing windows:

\begin{enumerate}
\item \textbf{Anatomical Regions (Table \ref{apptab:region}):} Evaluation of specific topographical zones (e.g., Upper Zone, Pulmonary Artery) to determine the spatial distribution and cardiovascular impact of the disease.
\item \textbf{Anatomical Structures (Table \ref{apptab:structure}):} Identification of the primary tissue types involved, distinguishing between "small airway disease" (Lung window) and vascular remodeling or right-heart strain (Mediastinal window).
\item \textbf{Pathological Signs (Table \ref{apptab:patho}):} Reporting of core COPD hallmarks, such as emphysematous destruction, hyperinflation, and mucus plugging.
\end{enumerate}

By decoupling the windows during annotation, we established a ``gold standard" for what information is uniquely available in each view. This allowed us to measure exactly how well our Knowledge Distillation Fusion architecture recovers mediastinal-specific structural expertise compared to standard multi-channel and supervised baselines.

\begin{table}[htbp]
\centering
\caption{Anatomical Region Terms for COPD Assessment}
\label{apptab:region}
\small
\begin{tabularx}{\textwidth}{l l l X}
\toprule
\textbf{English Term} & \textbf{Viewing Window} & \textbf{Relevant to COPD?} & \textbf{Clinical Note} \\ \midrule
None & N/A & No & Normal finding. \\
Bilateral & Lung / Mediastinal & Yes (High) & COPD is a systemic, diffuse disease affecting both lungs. \\
Hilum & Mediastinal & Yes & Site of large pulmonary arteries; enlargement suggests hypertension. \\
Upper Zone & Lung (HRCT) & Yes (High) & Classic site for smoking-related emphysema. \\
Middle Zone & Lung (HRCT) & Yes & Involved as disease progresses. \\
Lower Zone & Lung (HRCT) & Yes & Involved in panacinar emphysema (e.g., AAT deficiency). \\
L/R Lower Lobe & Lung (HRCT) & Yes & Common sites for mucus plugging and secondary infection. \\
L/R Base & Lung (HRCT) & Yes & Often shows flattened diaphragms due to hyperinflation. \\
Pulmonary Artery & Mediastinal & Yes (High) & Dilation ($>29$mm) is a major sign of COPD-related hypertension. \\
Right Atrium & Mediastinal & Yes (Advanced) & Enlargement indicates right-sided heart strain (Cor Pulmonale). \\
Right Ventricle & Mediastinal & Yes (Advanced) & Hypertrophy or dilation is a common COPD complication. \\
Left Ventricle & Mediastinal & No & Typically unaffected unless the patient has co-existing heart disease. \\ \bottomrule
\end{tabularx}
\end{table}

\begin{table}[htbp]
\centering
\caption{Anatomical Structure  Terms for COPD Assessment}
\label{apptab:structure}
\small
\begin{tabularx}{\textwidth}{l l l X}
\toprule
\textbf{English Term} & \textbf{Viewing Window} & \textbf{Relevant to COPD?} & \textbf{Clinical Note} \\ \midrule
Large-Medium Airways & Lung (HRCT) & Yes & Site of chronic inflammation/bronchitis. \\
Small Airways & Lung (HRCT) & Yes (Key) & COPD is primarily a ``small airway disease.'' \\
Parenchyma & Lung (HRCT) & Yes & Where emphysematous destruction occurs. \\
Bony Thorax & Bone Window & Yes & Changes shape (barrel chest) due to lung disease. \\
Right Heart & Mediastinal & Yes & COPD can lead to Cor Pulmonale (right heart failure). \\
Pulmonary Artery & Mediastinal & Yes & Enlargement indicates Pulmonary Hypertension from COPD. \\
Pulmonary Vessels & Mediastinal & Yes & Vascular pruning occurs as lung tissue is destroyed. \\
Pleura & Mediastinal & Indirectly & Usually secondary (e.g., pneumothorax from a burst bulla). \\
Interstitium & Lung (HRCT) & No & Usually refers to ``Restrictive'' diseases (like Fibrosis). \\
Left Heart & Mediastinal & No & Typically not directly affected by COPD. \\
None & N/A & No & Normal finding. \\ \bottomrule
\end{tabularx}
\end{table}

\begin{table}[htbp]
\centering
\caption{Abnormal Pathological Signs for COPD Assessment}
\label{apptab:patho}
\small
\begin{tabularx}{\textwidth}{l l l X}
\toprule
\textbf{English Term} & \textbf{Viewing Window} & \textbf{Relevant to COPD?} & \textbf{Clinical Note} \\ \midrule
Emphysema & Lung (HRCT) & Yes (Core) & One of the two main types of COPD. \\
Hyperinflation & Lung (HRCT) & Yes (Core) & Result of ``air trapping'' common in COPD. \\
Barrel Chest & Mediastinal/Bone & Yes (Core) & Physical sign of long-term hyperinflation. \\
Airway Wall Thickening & Lung (HRCT) & Yes (Core) & Sign of chronic bronchitis (the other COPD type). \\
Mucus Plugging & Lung (HRCT) & Yes & Common in the ``chronic bronchitis'' phenotype. \\
Bullae & Lung (HRCT) & Yes & Often a complication of advanced emphysema. \\
None & N/A & No & Normal finding. \\ \bottomrule
\end{tabularx}
\end{table}

\section{Further Analysis of Cross-Window Knowledge Distillation}
\label{app:mechanism}

To further investigate the source of the performance gains brought by cross-window knowledge distillation over supervised baselines, we performed a sample-level comparison between the two approaches on the COPD-CT-DF test set. The analysis is carried out from two complementary perspectives: a Venn-diagram analysis that quantifies how distillation corrects or deviates from supervised predictions, and a Grad-CAM visualization \citep{selvaraju2017grad} that exposes the differences in the spatial attention patterns of the two types of models.

\subsection{Venn-Diagram Analysis: Prediction Agreement and Correction}

\begin{figure*}[!htbp]
  \centering
  \begin{subfigure}[t]{0.48\linewidth}
    \centering
    \includegraphics[width=\linewidth]{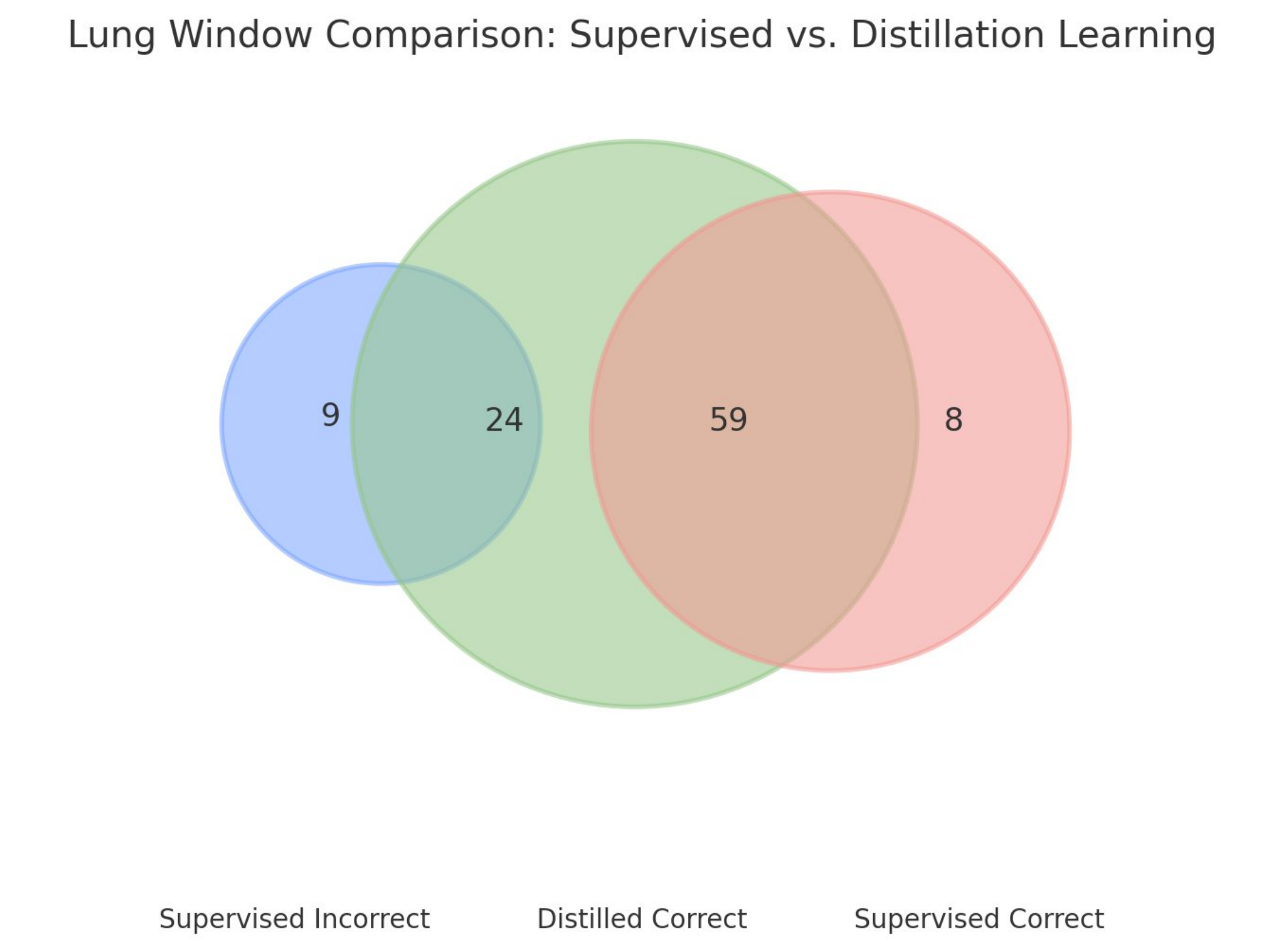}
    \caption{Lung window.}
    \label{fig:venn_lung}
  \end{subfigure}
  \hfill
  \begin{subfigure}[t]{0.48\linewidth}
    \centering
    \includegraphics[width=\linewidth]{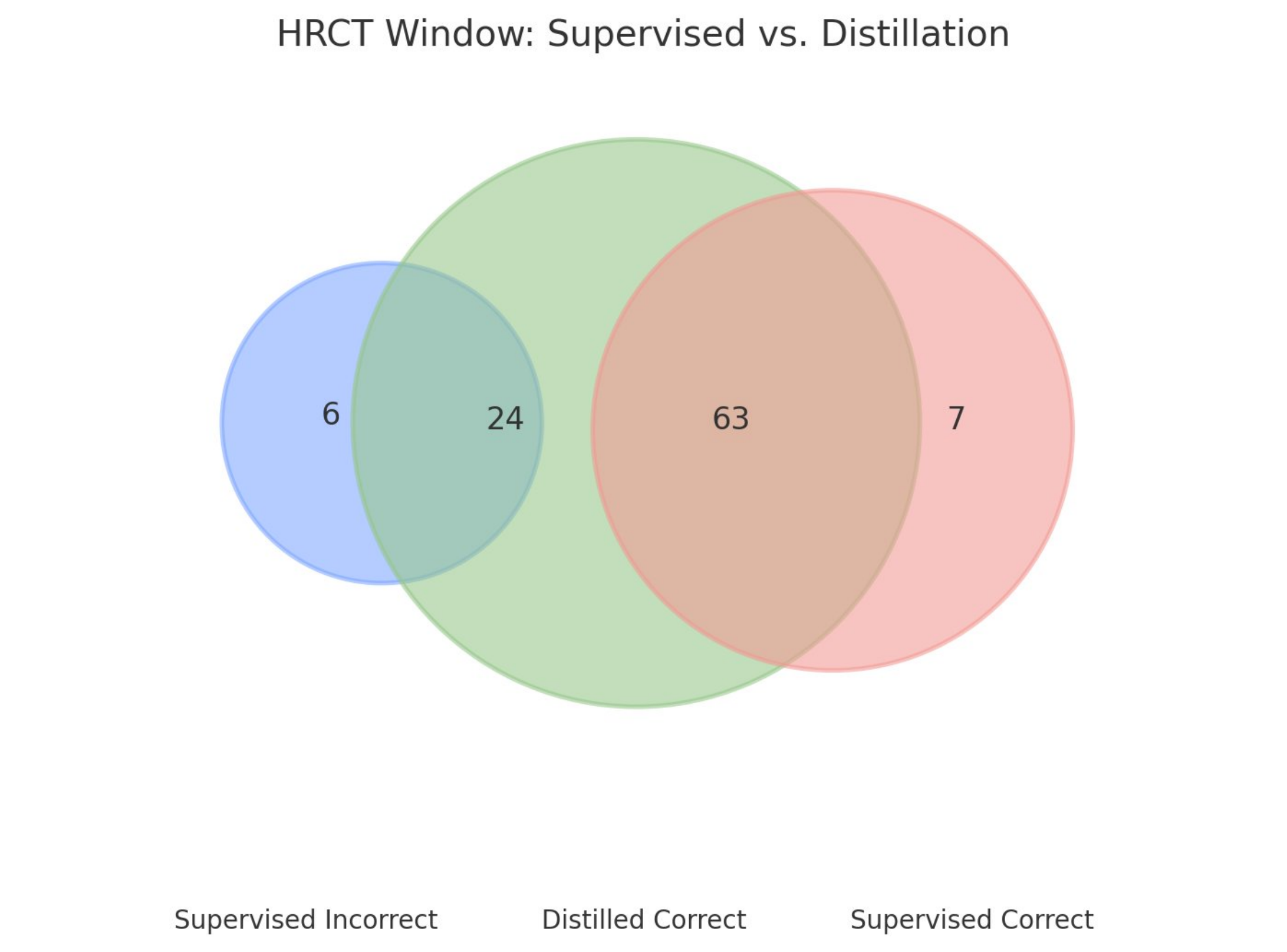}
    \caption{HRCT window.}
    \label{fig:venn_hrct}
  \end{subfigure}

  \vspace{0.6em}

  \begin{subfigure}[t]{0.48\linewidth}
    \centering
    \includegraphics[width=\linewidth]{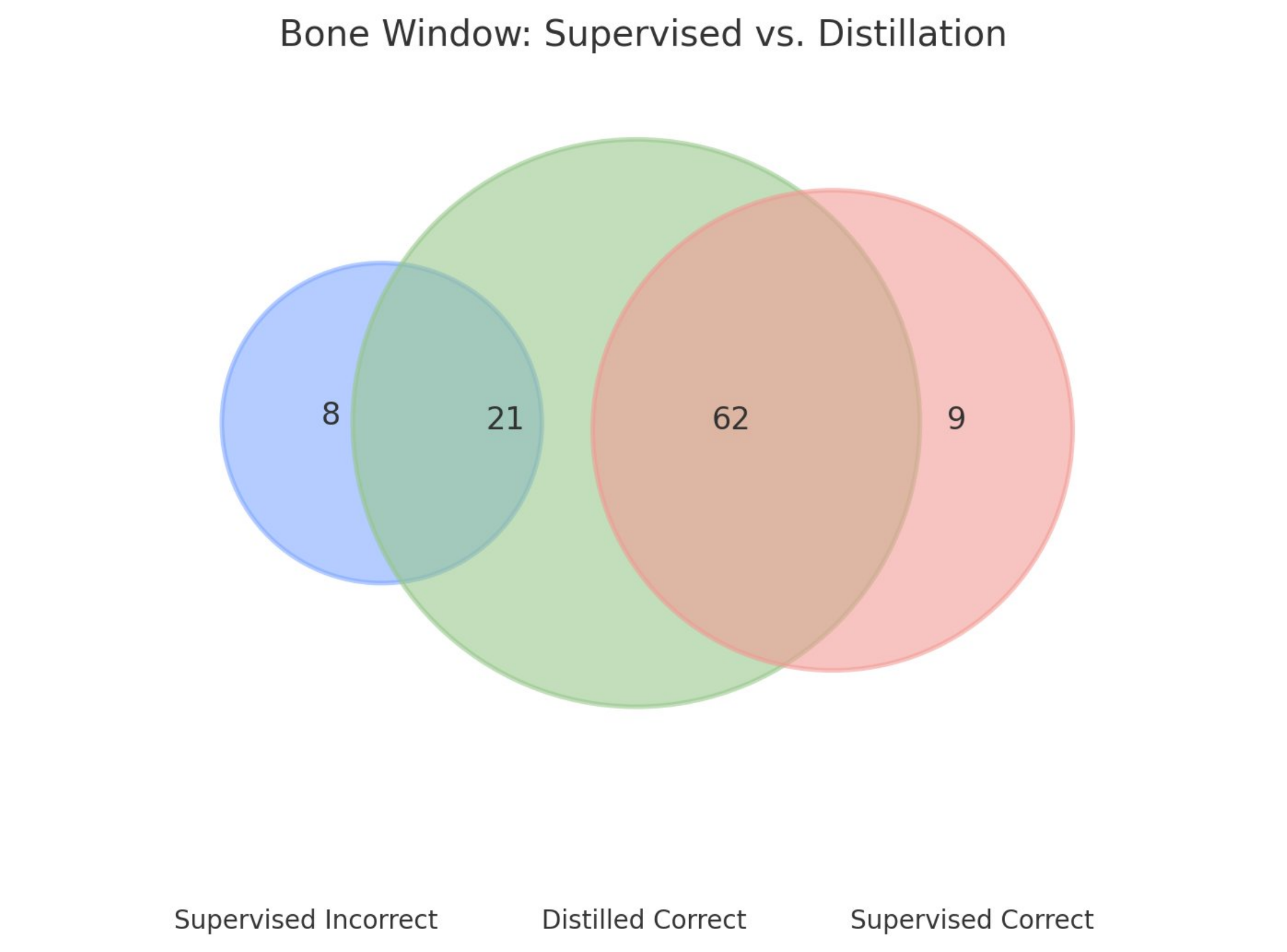}
    \caption{Bone window.}
    \label{fig:venn_bone}
  \end{subfigure}
  \hfill
  \begin{subfigure}[t]{0.48\linewidth}
    \centering
    \includegraphics[width=\linewidth]{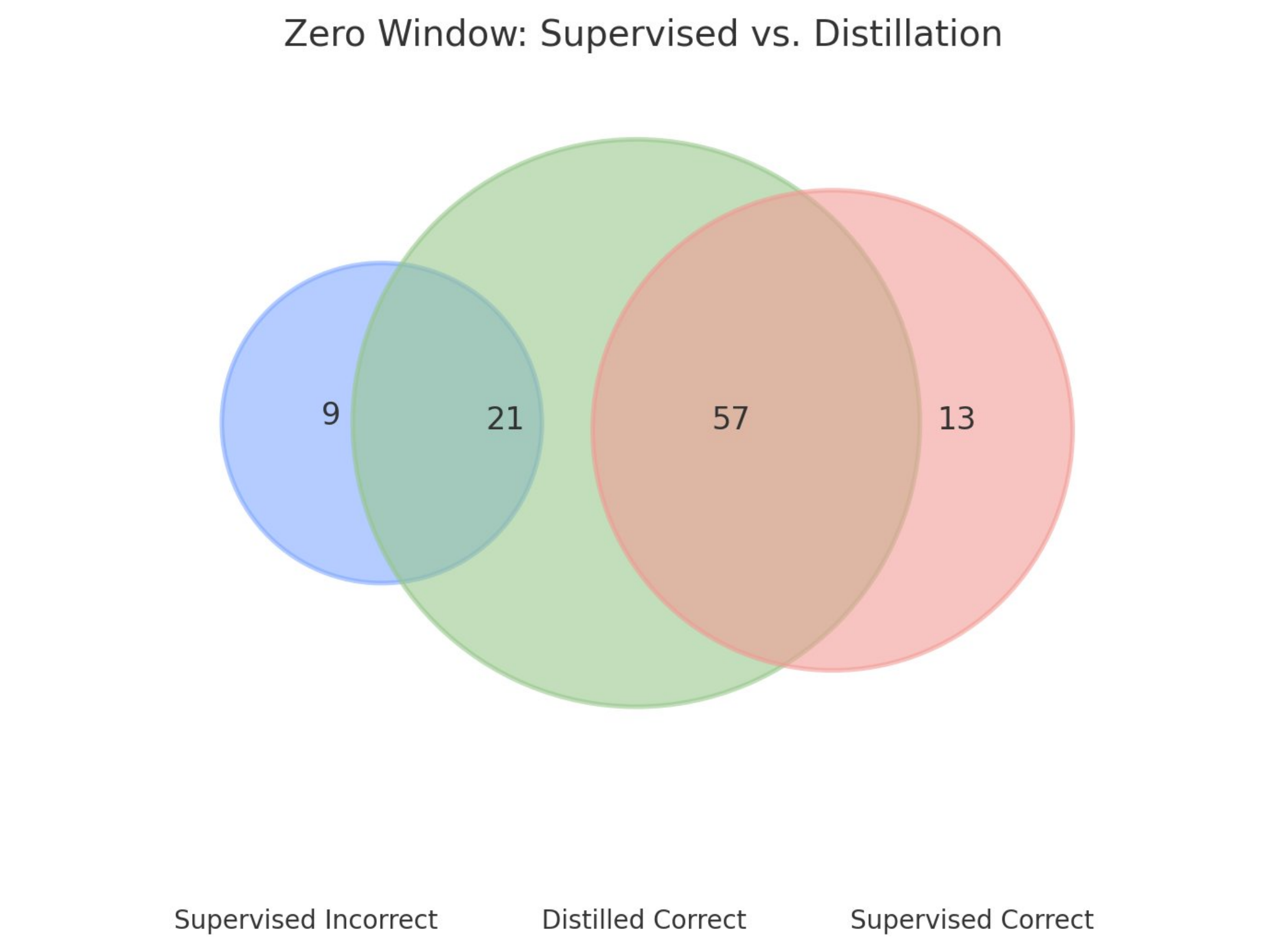}
    \caption{Zero window.}
    \label{fig:venn_zero}
  \end{subfigure}

  \caption{Per-sample prediction agreement between supervised learning and cross-window knowledge distillation on the four COPD-CT-DF student windows: (a) lung, (b) HRCT, (c) bone, and (d) zero. In each Venn diagram, the blue circle represents samples misclassified by the supervised model and the red circle represents samples correctly classified by the supervised model; the green circle represents samples correctly classified by the distilled model. The intersection of the blue and green circles thus indicates cases corrected by distillation (supervised-incorrect but distilled-correct), the intersection of the red and green circles indicates jointly correct predictions, and the portion of the red circle outside the green circle indicates cases that become incorrect after distillation.}
  \label{fig:distillation_analysis_app}
\end{figure*}

For each of the four student windows (zero, bone, HRCT, and lung), we drew a Venn diagram comparing the per-sample predictions of the supervised and distilled models (Figure~\ref{fig:distillation_analysis_app}). In every diagram, the left-hand region corresponds to cases that were misclassified under supervised learning but corrected after distillation; the central overlap corresponds to cases correctly classified by both models; and the right-hand region corresponds to cases that were correctly classified under supervised learning but then misclassified after distillation, i.e.\ new errors introduced by distillation.

The four Venn diagrams collectively reveal a consistent pattern: on every student window, distillation corrected a substantial number of previously misclassified cases while introducing far fewer new errors. Specifically, the number of corrected cases was 21 (zero window), 21 (bone window), 24 (HRCT window) and 24 (lung window); the number of jointly correct predictions ranged from 57 to 63; and the number of newly introduced errors ranged from 7 to 13. The pronounced asymmetry between the correction and new-error counts indicates that the gains brought by cross-window distillation do not reflect random fluctuations in the predictions but rather a directional improvement in discriminative capability.

\subsection{Grad-CAM Analysis: Changes in Attention Patterns}

\begin{figure*}[!htbp]
\centering
\includegraphics[width=\textwidth]{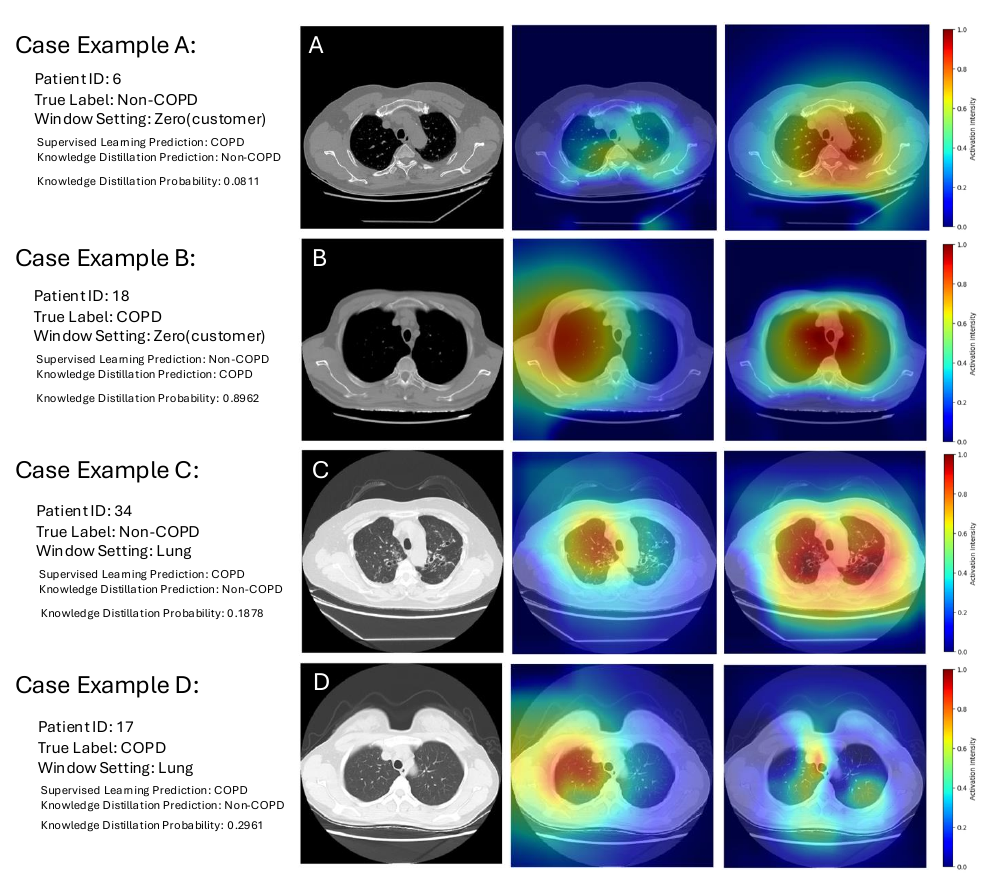}
\caption{Grad-CAM visualization comparing supervised and distilled models on four representative COPD-CT-DF cases. For each case, the three columns show the original CT slice, the Grad-CAM heatmap of the supervised model, and the Grad-CAM heatmap of the distilled model. Cases~A--C illustrate successful corrections of supervised mispredictions, while Case~D illustrates an ambiguous case in which distillation introduces a new error.}
\label{fig:gradcam_analysis_app}
\end{figure*}

To further understand how distillation reshapes the discriminative evidence used by the models, we used Grad-CAM to visualize the attention heatmaps of both models on selected cases, and randomly chose four representative examples (Figure~\ref{fig:gradcam_analysis_app}). The chosen cases cover the two principal types of changes introduced by distillation: on the one hand, representative examples in which false positives or false negatives were successfully corrected (Cases~A, B and~C); on the other hand, an ambiguous case in which distillation instead introduced a new error (Case~D), so as to present a complete picture of the distillation effect.

Overall, compared with the supervised models, the distilled models exhibited more focused and disease-relevant attention patterns. When supervised models spread their attention over non-disease-related regions such as the chest wall or mediastinal boundaries and thereby produced misclassifications, the distilled models tended to concentrate their attention on the pulmonary parenchyma and vascular structures; and on several COPD cases that the supervised models failed to identify, the distilled models formed a more complete response over the bilateral parenchymal changes. These visualizations are consistent with the correction trends observed in the Venn-diagram analysis and suggest that the gains from cross-window distillation are achieved through more accurate localization of discriminative features, rather than through a mere shift in prediction probabilities.

\subsection{Cross-Task Transfer Performance Across Metrics}
Table~\ref{tab:pe_fusion_results} reports the full per-metric performance of the supervised and the distilled ensembles when transferred from the RSNA PE dataset to the CTPA dataset under both transfer settings. Under \emph{direct transfer}, the distilled ensemble achieved an accuracy of 0.5938, a precision of 0.3404, an F1-score of 0.4507, a recall of 0.6667, and an AUC of 0.6603, compared with 0.5208, 0.3158, 0.4390, 0.7200, and 0.5944 for the supervised ensemble. Under \emph{fine-tuned transfer}, the distilled ensemble achieved an accuracy of 0.6776, a precision of 0.4718, an F1-score of 0.5724, a recall of 0.7276, and an AUC of 0.7481, compared with 0.6388, 0.4761, 0.4255, 0.3846, and 0.6264 for the supervised ensemble.

\begin{table*}[ht]
\centering
\caption{Comparison of models trained on the RSNA PE dataset and transferred to the CTPA dataset (two different types of transfer).}
\label{tab:pe_fusion_results}
\begin{tabular}{llcccccc}
\toprule
\textbf{Strategy} & \textbf{Method} & \textbf{Acc} & \textbf{Precision} & \textbf{F1} & \textbf{Recall} & \textbf{AUC} \\
\midrule
Direct Fusion
 & Supervised Learning    & 0.5208 & 0.3158 & 0.4390 & 0.7200 & 0.5944 \\
 & Knowledge Distillation & 0.5938 & 0.3404 & 0.4507 & 0.6667 & \textbf{0.6603} \\
\midrule
Fine-tuned Fusion
 & Supervised Learning    & 0.6388 & 0.4761 & 0.4255 & 0.3846 & 0.6264 \\
 & Knowledge Distillation & \textbf{0.6776} & \textbf{0.4718} & \textbf{0.5724} & \textbf{0.7276} & \textbf{0.7481} \\
\bottomrule
\end{tabular}
\end{table*}

\end{document}